\documentclass[journal]{IEEEtran}

\usepackage{caption}
\usepackage{subcaption}
\usepackage{makeidx}  
\usepackage{graphicx}
\usepackage{gensymb}
\usepackage{amssymb}  
\usepackage[ruled]{algorithm2e}
\usepackage{bm}
\usepackage{caption}
\usepackage{textcomp}
\usepackage{booktabs} 
\usepackage{amsmath}
\usepackage{multirow}
\usepackage{url}
\usepackage{nicefrac}
\usepackage{hyperref}
\usepackage{color}

\usepackage[flushmargin]{footmisc}

\usepackage[linewidth=1pt,framemethod=TikZ]{mdframed}
\usepackage{float}

\begin{document}

\title{Explainable Anatomical Shape Analysis through Deep Hierarchical Generative Models}

\author{Carlo~Biffi, Juan~J.~Cerrolaza, Giacomo~Tarroni, Wenjia~Bai,  Antonio~de~Marvao, Ozan~Oktay, \\ Christian Ledig, Loic~Le~Folgoc, Konstantinos~Kamnitsas, Georgia~Doumou, Jinming~Duan,\\ Sanjay~K.~Prasad, Stuart~A.~Cook, Declan~P.~O'Regan, and Daniel~Rueckert
\thanks{Copyright (c) 2020 IEEE.  Personal use of this material is permitted.  Permission from IEEE must be obtained for all other uses, in any current or future media, including reprinting/republishing this material for advertising or promotional purposes, creating new collective works, for resale or redistribution to servers or lists, or reuse of any copyrighted component of this work in other works.}
\thanks{The research was supported by grants from the British Heart Foundation (NH/17/1/32725, RE/13/4/30184), Academy of Medical Sciences (SGL015/1006) and the National Institute for Health Research Biomedical Research Centre based at Imperial College Healthcare NHS Trust and Imperial College London. \emph{(Corresponding author: Carlo Biffi, e-mail: c.biffi15@imperial.ac.uk)}}%
\thanks{C. Biffi, J. J. Cerrolaza, G. Tarroni, W. Bai, O. Oktay, C. Ledig, L. Le Folgoc, K. Kamnitsas, J. Duan and D. Rueckert are with the Department of Computing, Imperial College London. S. K. Prasad is with the National Heart and Lung Institute, Imperial College London. A. de Marvao, G. Doumou, D. P. O'Regan and S. A. Cook are with the MRC London Institute of Medical Sciences, Faculty of Medicine, Imperial College London.}
}

\maketitle

\begin{abstract}
Quantification of anatomical shape changes currently relies on scalar global indexes which are largely insensitive to regional or asymmetric modifications. Accurate assessment of pathology-driven anatomical remodeling is a crucial step for the diagnosis and treatment of many conditions. Deep learning approaches have recently achieved wide success in the analysis of medical images, but they lack interpretability in the feature extraction and decision processes. In this work, we propose a new interpretable deep learning model for shape analysis. In particular, we exploit deep generative networks to model a population of anatomical segmentations through a hierarchy of conditional latent variables. At the highest level of this hierarchy, a two-dimensional latent space is simultaneously optimised to discriminate distinct clinical conditions, enabling the direct visualisation of the classification space. Moreover, the anatomical variability encoded by this discriminative latent space can be visualised in the segmentation space thanks to the generative properties of the model, making the classification task transparent. This approach yielded high accuracy in the categorisation of healthy and remodelled left ventricles when tested on unseen segmentations from our own multi-centre dataset as well as in an external validation set, and on hippocampi from healthy controls and patients with Alzheimer's disease when tested on ADNI data. More importantly, it enabled the visualisation in three-dimensions of both global and regional anatomical features which better discriminate between the conditions under exam. The proposed approach scales effectively to large populations, facilitating high-throughput analysis of normal anatomy and pathology in large-scale studies of volumetric imaging.
\end{abstract}

\begin{IEEEkeywords}
Shape Analysis, Explainable Deep Learning, Generative Modeling, MRI.
\end{IEEEkeywords}

\IEEEpeerreviewmaketitle

\section{Introduction}
\IEEEPARstart{T}{he} quantification of anatomical changes and their relationship with disease is a fundamental task in medical image analysis, ultimately leading to new clinical insights and enhanced risk assessment and treatment. Recent improvements in the medical image analysis field have been characterised by an increase of large-scale population-based initiatives \cite{fonseca2011cardiac}, \cite{attar2019quantitative}, \cite{mueller2005alzheimer} together with development of automated segmentation pipelines of anatomical structures \cite{nogovitsyn2019testing}, \cite{ledig2018structural}, which recently achieved human-level performance \cite{bai2018automated}. In this context, the development of novel data-driven processing tools to enable quantitative assessment of the differences between normal anatomy and pathology has now received significant interest \cite{bruse2017detecting},  \cite{triposkiadis2019continuous}, \cite{shen2012detecting}.

Alterations in shape and structure of an organ associated with an underlying pathology, here defined as pathological remodelling, are of particular interest for the classification and risk-stratification of patients. Hypertrophic cardiomyopathy (HCM) is a cardiac disease defined by the presence of left ventricular (LV) hypertrophy that cannot be solely explained by abnormal loading conditions \cite{yancy20132013}. In HCM, hypertrophy manifests in complex regional patterns not readily quantifiable using volumetric indices \cite{captur2016embryological}. Similarly, atrophic changes in the hippocampus are considered as relevant biomarkers for the diagnosis and prediction of Alzheimer's disease (AD), and proved to differently affect distinct local areas of the hippocampal shape \cite{frisoni2008mapping}, \cite{shen2012detecting}. For most human organs, the gold-standard imaging technique to assess structural shape changes is magnetic resonance (MR) which enables imaging at high-resolution and in three-dimensions (3D) \cite{elliott2014esc}, \cite{ledig2018structural}. Despite the advances in MR imaging, classification and risk-stratification of patients still rely on scalar indexes describing pathological remodeling (e.g. left ventricular mass or hippocampal volume), which neglect regional or asymmetric effects that occur during pathology whose quantification could improve early detection and risk stratification \cite{triposkiadis2019continuous}, \cite{elliott2014esc}, \cite{frisoni2008mapping}, \cite{shen2012detecting}.


Machine learning approaches have achieved outstanding results in the medical image analysis domain, such as in the discrimination of physiological versus pathological hypertrophy patterns from multiple manually-derived cardiac indices \cite{narula2016machine}, between patients with dilated cardiomyopathy patients and controls \cite{puyol2019regional} and of patients with AD and mild cognitive impairment patients as well as healthy controls \cite{basaia2019automated}. In particular, deep learning methods proved to be powerful features extractors for the classification of clinical conditions from medical images\cite{litjens2017survey}, \cite{bernard2018deep}. Despite their tremendous success, however, a major drawback is their lack of interpretability,  which currently hampers their translation to clinical practice. In fact, the physiological reason that drives the classification result is often as important as the classification result itself \cite{bernard2018deep}.

In this work, we propose a new deep learning approach to learn a hierarchy of conditional latent variables that (1) models a population of anatomical segmentations of interest, (2) enables the classification of distinct clinical conditions by using the highest level of the hierarchy and (3) whose anatomical effect can be visualised and quantified in the original segmentation space. These contributions are achieved by specialising the highest level of a deep hierarchical generative model for the classification of distinct clinical conditions. As a consequence, thanks to the generative properties of the model, distinct segmentations corresponding to different values of the highest level can be generated, making the classification model interpretable. In addition, by constraining the highest level to be two-dimensional, the feature space in which the classification is performed can also be directly visualised. Therefore, our approach consists in an automated data-driven tool which enables the detailed analysis of the pathological remodelling patterns associated with a large number of clinical conditions.

\section{Related Work}
An autoencoder is a non-linear dimensionality reduction technique which learns a compact feature representation of the input data by encoding it into and decoding it from a low-dimensional feature vector. Deep autoencoder-based architectures have achieved wide success in computer vision applications as an extension of PCA-based approaches, including feature learning of 3D objects \cite{fang20153d}. Autoencoder-based models have also been used to learn compact representations of medical images \cite{litjens2017survey}. Relevant to this work, Oktay \emph{et al.}~\cite{oktay2018anatomically} showed how autoencoder-derived features of LV segmentations outperform PCA features in the classification of healthy subjects versus dilated cardiomyopathy and HCM patients.

Deep generative models have demonstrated great performance in learning data distributions over a low-dimensional set of latent variables and in generating new unseen samples, which is not possible with standard autoencoder models. Within this class of models, variational autoencoder (VAE) models \cite{kingma2013auto} learn a continuous latent representation by enforcing it to behave according to a predefined distribution. VAEs have been successful at learning the latent space representing deforming 3D shapes for a variety of applications, including shape space embedding and generation, outperforming state-of-the-art methods \cite{nash2017shape}, \cite{tan2018variational}. In the medical imaging domain, VAEs have been exploited to approximate the distribution and likelihood of previously unseen MR images \cite{tezcan2018mr}, to learn a low-dimensional manifold of 3D fetal skull segmentations \cite{cerrolaza20183d} and to learn a low-dimensional probabilistic deformation model for cardiac image registration \cite{krebs2019learning}.

Hierarchical VAEs are a class of generative models that decompose the input data into a hierarchical representation \cite{rezende2014stochastic}, \cite{sonderby2016ladder}. Although highly flexible, these models have been traditionally difficult to optimise, especially in the training of their higher levels, as often their lowest layer alone can contain enough information to reconstruct the data distribution, and the other levels are ignored. In this work, we focus on the ladder VAE (LVAE) framework \cite{sonderby2016ladder}, which was shown to be capable of learning a deeper and more distributed latent representation by combining the approximate likelihood and the data-driven prior latent distribution at each level of the generative model.

In hippocampus shape analysis, Shakeri \emph{et al.}~\cite{shakeri2016deep} employed a VAE model to learn a low-dimensional representation of co-registered hippocampus meshes, which was employed in conjuction with a multi-layered perceptron (MLP) to classify healthy subjects from AD patients. The network input consisted of mesh vertices coordinates, and the representation was learned through two fully connected layers. Similarly, in our preliminary work \cite{biffi2018learning} we modified the 3D convolutional VAE framework in order to learn a low-dimensional latent representation of 3D LV segmentations, which was not only able to encode the 3D segmentations manifold, but also to discriminate different conditions by performing the classification task in the latent space. In the same work, we proposed a latent space navigation method to explore the anatomical variability encoded by the learned latent space. This consisted in iteratively modifying the latent representation of a segmentation obtained from an healthy subject along the direction that maximized its probability to be classified as pathological. By decoding the different latent representations in the original space of the segmentations, our technique allowed  the visualisation of the anatomical changes caused by this transformation. 

The following limitations characterize our preliminary work: 1) The learned VAE latent space not only encoded the factors of variation that most discriminate between classes, but also all the other factors of variation that regulate shape appearance. The latent space navigation was thus a necessary step to attempt the offline estimation of the variations linked to the pathological remodeling. In this work, we aim at automatically learning a latent space that encodes only these changes. 2) Our previous work required an additional offline dimensionality reduction technique to visualize in two dimensions the clustering obtained in the VAE latent space, which would however not reflect the real distribution of the shapes in the learned latent space. In this work, we aim at directly learning this two-dimensional latent space. 3) The latent space navigation method proposed in our previous work could only obtain subject-specific paths (with no obvious navigation stopping criteria). In this work, we aim at providing a means to extract the more clinically appealing population-based inferences.

In the later work of Bello \emph{et al.} \cite{bello2019deep}, a supervised denoising autoencoder was used to learn a latent code representation of right ventricular contraction patterns and, at the same time, to perform survival prediction. Not being a generative model, the effect of task-specific features learned by the proposed model could not be visualised, making the prediction task not explainable and population based inferences difficult to obtain. In addition, an additional offline dimensionality reduction step was also required to visualise in two-dimensions the distribution of different groups of subjects.

\subsubsection*{Contributions}
In this paper, we aim to extend our preliminary work \cite{biffi2018learning} on classification and visualisation of discriminative features by employing LVAEs, with the aim of assisting clinicians in quantifying the morphological changes related to disease, and in order to develop medical image classifiers that can visualise the morphological features driving the classification result. The main contributions of this work can be described as follows:

\begin{itemize}
\item We demonstrate that an interpretable classifier of anatomical shapes can be developed by performing a classification task of interest in the highest level of a LVAE model. In this way, the latent variables of this level automatically encode the most discriminative features for the task under exam, while the other subsequent levels model the remaining factors of anatomical variation in the data.

\item We show that the LVAE highest latent space can be assumed to be two-dimensional so that the classification space can be directly visualised without further offline dimensionality reduction steps. Furthermore, we demonstrate how the anatomical variability encoded by this latent space can be visualised in the original space of the segmentations thanks to the generative properties of the model, enabling the visualisation of the anatomical effect of the most discriminative features between different conditions. 

\item  We demonstrate how the proposed LVAE-based method achieves high classification accuracy of HCM versus healthy 3D LV segmentations and of AD versus healthy controls 3D hippocampal segmentations. More importantly, we show how the model captures and enables the easy visualisation of the most discriminative features between the conditions under exam. Finally, we show that the learned hierarchical representations provide higher reconstruction accuracy compared to single-latent-space VAEs.

\item While hierarchical VAEs have been mainly evaluated  on benchmark datasets, here we successfully apply them on two real-world 3D medical imaging datasets. We show insights on the model functioning and optimal training, and we make the implementation of proposed method publicly available\footnote{ https://github.com/UK-Digital-Heart-Project/lvae\_mlp\\ DOI 10.5281/zenodo.3247898}.
\end{itemize}

\section{Methods}
\begin{figure}[!t]
\centering
\includegraphics[width=0.95\columnwidth]{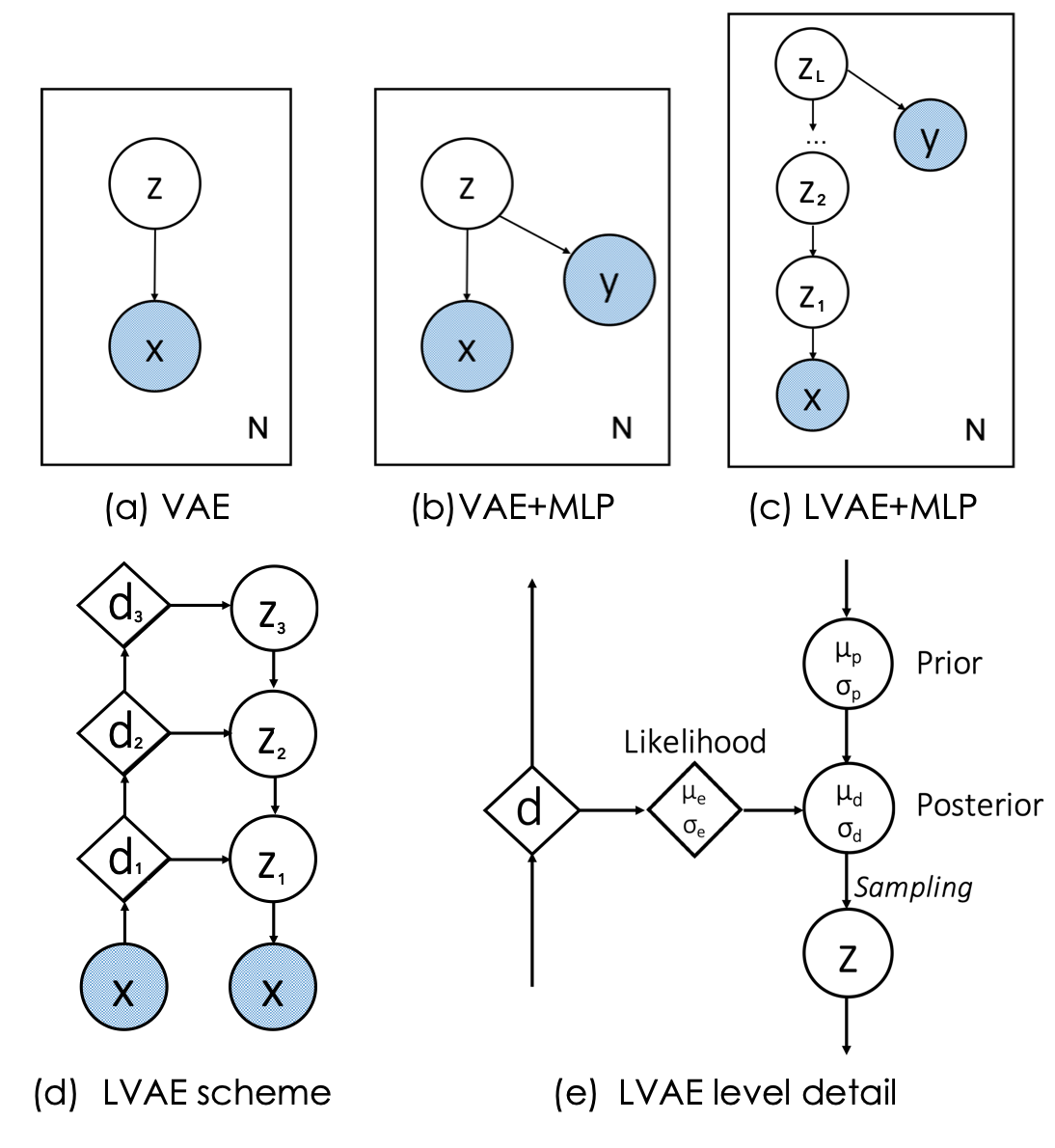}
\caption{Graphical models of a standard VAE (a), of our previously proposed method \cite{biffi2018learning} (b) and the new LVAE-based approach (c). $x$ represents and anatomical segmentation, $y$ the disease class label and $z$ the latent variables to learn. Schematic representation of a three-level LVAE (d) and of the flow of information (e). Circles represent stochastic variables, diamonds represent deterministic variables. Variables in light blue represent the inputs of the network.}
\label{fig:modelG}
\end{figure}

This section is organised as follows. First, in subsection A and B, we summarise the theoretical foundations of the proposed method. Second, in subsection C, we describe our modifications to the original VAE and LVAE frameworks towards explainable shape analysis (graphical models in Fig. \ref{fig:modelG}). Then, in subsection D, we describe the datasets used in this work for the classification of healthy subjects versus HCM patients and of heathy controls versus AD patients. Finally, in subsection E, we provide a detailed description of the LVAE models used in this work (model summary in Fig. \ref{fig:LVAEarchi} for the cardiac application). 

\subsection{Variational Autoencoder (VAE)}
\label{secVAE}
Given a training set of $N$ anatomical segmentations $X=\{ x_j , \; j=1,...N \}$ of a structure of interest from a population $S$, a VAE \cite{kingma2013auto} is a probabilistic generative model that aims at learning the distribution $p_{\theta}(x)$ of the population of segmentations $x \in S$ under study. The distribution $p_{\theta}(x)$ is learned from the data by using a model of latent variables $z \in \mathcal{R}^p$, where $p \ll d$ and $d$ is the number of pixels/voxels in a segmentation $x \in S$. The VAE graphical model is depicted in Fig. \ref{fig:modelG} (a) and the generative model is defined as
\begin{equation}
p_{\theta}(x) = \int_z p_{\theta}(x,z) dz = \int_z p_{\theta}(x|z) p_{\theta}(z) dz
\label{uno}
\end{equation}

where $p_{\theta}(z)$ is the prior distribution over the variables $z$, $p_{\theta}(x|z)$ is the generative (or decoder) network and $\theta$ are the learnable parameters of the model. However, directly optimising $\log(p_{\theta}(x))$ for the $N$ segmentations of the training set $X$ is computationally infeasible, as it requires to compute the integral in Eq. \ref{uno} over all the $z$ values. The VAE framework addresses this issue by introducing a variational distribution $q_{\phi}(z|x)$  to approximate the posterior distribution of the latent variables $z$, $p_{\theta}(z|x)$. After applying Bayes' rule and rearranging \cite{doersch2016tutorial}, the following equation can be derived

\begin{multline}
log(p(x)) - KL[q_{\phi}(z|x) || p_{\theta}(z|x)]  = \\ E_{q_{\phi}(z|x)}[\log(p_{\theta}(x|z))] - KL[q_{\phi}(z|x) || p_{\theta}(z)]
\label{due}
\end{multline}

where KL is the Kullback-Leibler (KL) divergence. By assuming that the $q_{\phi}(z|x)$ is modeled with an high capacity function, the right-hand side of Eq. (2) becomes a lower bound for $\log(p_{\theta}(x))$ and can be optimized via stochastic gradient descent. The first term in the lower bound represents a reconstruction loss, i.e. how accurate is the generative model $p_{\theta}(x)$ in the reconstruction of the segmentation $x$ from the latent space values $z$ using the generative (or decoder) network $p_{\theta}(x|z)$. The second term is a regularization term that makes $q_{\phi}(z|x)$ match with its prior distribution $p_{\theta}(z)$ on the latent variables $z$.

\subsection{Ladder Variational Autoencoder (LVAE)}
A Ladder VAE (LVAE) \cite{sonderby2016ladder} is a hierarchical latent variable model that employs a hierarchy of $i=1,...,L$ conditional latent variables in the generative model and it is schematised in Fig. \ref{fig:modelG} (d). The total prior distribution $p_{\theta}(z)$ of this model is factorised as:
\begin{equation}
p_{\theta}(z) = p_{\theta}(z_L) \prod_{i=1}^{L-1} p_{\theta}(z_i|z_{i+1})
\end{equation}
\begin{equation}
p_{\theta}(z_i|z_{i+1}) = \mathcal{N}(z_i|\mu_{p,i}(z_{i+1}), \sigma^2_{p,i}(z_{i+1}))   \;\;\; \forall i < L
\label{eq4}
\end{equation}
\begin{equation}
p_{\theta}(z_L) = \mathcal{N}(z_L|0,1)
\label{eq5}
\end{equation}
where the highest latent space ($i=L$) has a prior distribution $p_{\theta}(z_L)$ which is typically assumed to be a Gaussian distribution with $\mu_{p,L}=0$ and $\sigma_{p,L}^2=1$ (Eq. \ref{eq5}), while the other levels in the hierarchy have their prior values of $\mu_{p,i}$ and $\sigma_{p,i}^2$ that conditionally depend on the upper levels of the ladder (Eq. \ref{eq4}).

The LVAE inference model also differs from a standard VAE. In particular, each layer $i$ in the hierarchy of the latent variables is conditioned on the previous stochastic layers and the total inference model $q_{\phi}(z|x)$ is specified by the following fully factorised Gaussian distribution:
\begin{equation}
q_{\phi}(z|x) = q_{\phi}(z_1|x)  \prod_{i=1}^{L-1} q_{\phi}(z_{i+1}|z_{i})
\label{eqFac}
\end{equation}
\begin{equation}
q_{\phi}(z_i|\cdot) = \mathcal{N}(z_i|\mu_{d,i},\sigma^2_{d,i}) 
\end{equation}
In contrast with standard hierarchical VAEs \cite{rezende2014stochastic}, where the inference $q_{\phi}(z|x)$ and prior distributions $p_{\theta}(z)$ are computed separately with no explicit sharing of information, the LVAE framework introduces a new inference mechanism. As shown in Fig. \ref{fig:modelG} (e), at each level $i$, an approximate likelihood estimation $\mu_{e,i}$ and $\sigma^2_{e,i}$ of its latent Gaussian distribution parameters is obtained from the encoder branch. This likelihood estimation is combined with the prior estimates $\mu_{p,i}$ and $\sigma^2_{p,i}$ obtained from the generative branch to produce a posterior estimation $\mu_{d,i}$ and $\sigma^2_{d,i}$ of the latent Gaussian distribution at that level $i$. In particular, this sharing mechanism between the inference (encoder) and generative (decoder) branches is performed at each level $i \ne L$ through a precision-weighted combination of the form:

\begin{equation}
\sigma_{d,i}^2 = \frac{1}{\sigma_{e,i}^{-2} + \sigma_{p,i}^{-2} } \; \; \; \; \; \; \; \;
\mu_{d,i} = \frac{\mu_{e,i} \sigma_{e,i}^{-2} + \mu_{p,i} \sigma_{p,i}^{-2}}{\sigma_{e,i}^{-2} + \sigma_{p,i}^{-2} }
\label{eqFac2}
\end{equation}

while $\mu_{d,L} = \mu_{e,L}$ and $\sigma^2_{d,L} = \sigma^2_{e,L}$. This  combination enables to build a data-dependent posterior distribution at each level, $\mathcal{N}(\mu_{d,i},\sigma^2_{d,i})$, that is both a function of the values assumed in the higher levels of the generative model and of the inference information derived of the subsequent (lower) levels. The loss function of the LVAE is the same of a VAE (Eq. \ref{due}) with the only difference that the number of KL divergence terms is equal to the number of levels $L$ in the ladder. These KL divergence terms force the learned prior and posterior distributions at each level to be as close as possible. The sharing of information between the encoder and decoder through Eq. \ref{eqFac2} promotes the learning of a data-dependent prior distribution better suited for the dataset to be modelled. Moreover, this provides a better and more stable training procedure as the inference (encoder) branch iteratively corrects the generative distribution, instead of learning the posterior and prior values separately \cite{sonderby2016ladder}.\\
The full LVAE generative model has therefore the following formulation:
\begin{equation}
p_{\theta}(x) = \int_z p_{\theta}(x|z_1) \; p_{\theta}(z_L) \prod_{i=1}^{L-1} p_{\theta}(z_i|z_{i+1}) \; dz
\label{lvaeGM}
\end{equation}

\subsection{LVAE for Interpretable Shape Analysis}
In our previous work \cite{biffi2018learning}, we proposed a modification of the standard VAE framework presented in Section \ref{secVAE} to include a classification network $p(y|z)$ able to predict the disease class label $y$ associated with a segmentation $x$ by using its latent representation $z$ (the corresponding graphical model is shown in Fig. \ref{fig:modelG} (b)). In this work, we hypothesise that such modification can be extended to the LVAE framework by connecting a MLP $p(y|z_L)$, which classifies the disease status $y$ of an input segmentation $x$, to the highest latent space $z_L$ (graphical model in Fig. \ref{fig:modelG} (c)). By training the LVAE+MLP architecture end-to-end we aim at obtaining a very low-dimensional latent space $z_L$ which encodes the most discriminative features for the classification task under study, while the other latent spaces will encode all the other factors of variation needed to reconstruct the input segmentations $x$. This has two main advantages: 1) template shapes for each disease class can be obtained by sampling from the learned distributions in a top-down fashion (starting from the highest level in the hierarchy $p(z_L|y)$ and subsequently from every prior $p_{\theta}(z_{i}|z_{i+1})$). The posterior $p(z_L|y)$ can be estimated by kernel density estimation and, since $z_L$ is typically very low-dimensional, this estimation is straightforward;  2) if the latent space $z_L$ is designed to be 2D or 3D, the distributions $p(z_L|y)$ in the classification space can be directly visualised without the need of further offline dimensionality reduction techniques required in previous works \cite{biffi2018learning}, \cite{bello2019deep}.

\subsection{Datasets}
 \textbf{Cardiac Dataset} A multi-centre cohort consisting of 686 HCM patients and 679 healthy volunteers was considered for this work. All subjects underwent cardiac phenotyping at a 1.5-T on Siemens (Erlangen, Germany) or Philips (Best, Netherlands) system using a standard cardiac MR protocol. HCM patients were confirmed with reference to established diagnostic criteria \cite{elliott2014esc}. 
 LV short-axis cine images were acquired with a balanced steady-state free-precession sequence. The end-diastolic (ED) and end-systolic (ES) phases were automatically segmented using a previously published and extensively validated cardiac multi-atlas segmentation framework \cite{bai2015bi}. As a first post-processing step, the obtained LV short-axis stack segmentations were upsampled using a multi-atlas label fusion approach. For each segmentation, twenty manually annotated high-resolution atlases at ED and ES were warped to the subject space using free-form non-rigid registration and fused with majority vote, leading to an upsampled high-resolution segmentation ($2 mm$ x $2  mm$ x $2 mm$) \cite{rueckert1999nonrigid}. In a second step, all segmentations were aligned onto the same reference space at ED by means of landmark-based and subsequent intensity-based rigid registration to remove pose variations. After extracting the LV myocardium label, each segmentation was cropped and padded to $[x=80, y=80, z=80, t=1]$ dimensions using a bounding box positioned at the centre of the LV ED myocardium. This latter operation guarantees shapes to maintain their alignment after cropping. Finally, all segmentations underwent manual quality control in order to discard scans with strong inter-slice motion or insufficient LV coverage, resulting in 436 HCM patients and 451 healthy volunteers that were used for the final analysis (population characteristics and standard CMR metrics are reported at Supplementary Data 6). As an additional external testing dataset, ED and ES segmentations from 20 healthy volunteers and 20 HCMs from the ACDC MICCAI’17 challenge training dataset\footnote{https://www.creatis.insa-lyon.fr/Challenge/acdc/} were also used (after undergoing pre-processing using the same high-resolution upsampling pipeline explained above).
 
 \begin{figure*}[!t]
\centering
\includegraphics[scale=0.22]{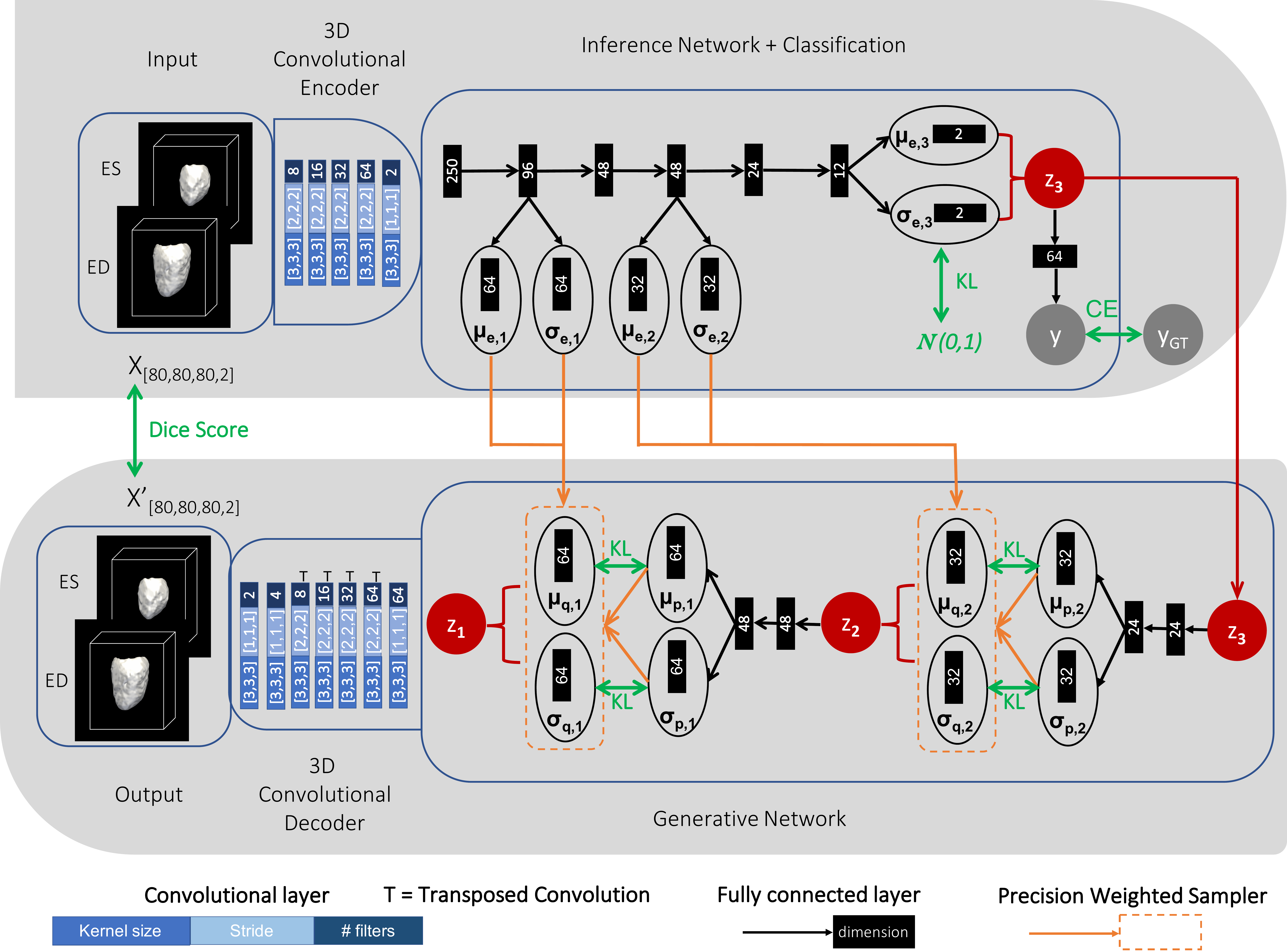}
\caption{Detailed scheme of the LVAE+MLP architecture adopted in this work for the cardiac application. Top: encoder model; Bottom: decoder model. At testing, segmentations class scores $y$ are computed with $z_3=\mu_{e,3}$. The green arrows indicate the loss function terms used to train the network.}
\label{fig:LVAEarchi}
\end{figure*}
 
\textbf{Brain Dataset}
A total of 726 3D left and right hippocampus segmentations of healthy controls (HC, $N=404$, 202 males, median age 74.2 [min$=$59.8;max$=$89.6]) and Alzheimer's disease subjects (AD, $N=322$, 177 males, median age 75.8 [min$=$55.1;max$=$91.4])  from a publicly available repository were analysed in this work \cite{ledig2018structural}. The segmentations were obtained from baseline T1-weighted (T1w) MR brain images from the ADNI-1/-GO/-2 cohorts using a multi-atlas label propagation method with expectation-maximisation based refinement (MALPEM) \cite{ledig2018structural}. Images were automatically segmented individually and no additional pre-processing was performed. All segmentations were rigidly registered to the MNI standard reference space using nearest neighbour interpolation. Shape-based interpolation was applied to upsample each segmentation to $0.75 mm$ x $0.75  mm$ x $0.75 mm$ resolution. Finally, each segmentation was cropped and padded using a bounding box positioned at its centre to obtain 3D segmentations of dimension $[x=60, y=60, z=60, t=1]$ for both the left and right hippocampus. Moreover, a 3D high-resolution left and right hippocampus template segmentation was obtained by averaging the upsampled and rigidly registered healthy controls segmentations. By thresholding the template probabilistic segmentation, a template triangular mesh was extracted using marching cubes algorithm which will be used in this work for results visualisation.  

\subsection{Application to Pathological Remodelling - LVAE+MLP model details}
 A detailed scheme of the three-level ($L=3$) LVAE+MLP architecture employed in this work for the classification of HCM patients versus healthy subjects is summarised in Fig. \ref{fig:LVAEarchi}, while the corresponding architecture for the classification of healthy controls versus AD patients is reported in Supplementary Materials 5. For the sake of display clarity the model scheme has been split into two rows: the encoder (inference) branch is shown at the top while the decoder (generative) branch is depicted at the bottom, and the two branches are connected by the latent space $z_3$. In the cardiac application, the input of the encoder branch are the 3D LV segmentations at ED and ES for each subject under study, which are presented as a two-channel input (top-left of Fig. \ref{fig:LVAEarchi}). A 3D convolutional encoder compresses them into a 250-dimensional embedding through a series of 3D convolutional layers with stride 2. This embedding is used then as input of a deterministic inference network, which computes the likelihood estimates $\mu_{e,i}$ and $\sigma_{e,i}$ for each level $i$ of the hierarchy of latent variables. These estimates are derived by manipulating the input through a series of fully connected layers (black arrows), which are all followed by batch normalisation and \emph{elu} non-linearity with the only exception of the layers computing $\mu_{e,i}$ and $\sigma_{e,i}$. At the highest latent space ($i=3$ in this case), a shallow MLP (2 layers) is attached to learn $p(y|z_3)$, i.e. to predict the class (HCM or healthy) label $y$ corresponding to the input segmentation $x$ by just using its latent variable values $z_3$. \emph{ReLu} was used as non-linearity after the first layer.  The latent variable values $z_3$ are sampled during training from $\mathcal{N}(\mu_{d,3},\sigma_{d,3}^2)$ where $\mu_{d,3} = \mu_{e,3}$ and $\sigma_{d,3} = \sigma_{e,3}$ and they are also the starting point of the generative process (bottom-right of Fig. \ref{fig:LVAEarchi}). At each level $i$ of the generative (decoder) network, the prior distribution terms are computed by modifying the values of the previous latent space $z_{i+1}$ through a fully connected layer followed by batch normalization and \emph{elu} non-linearity and by a second fully connected layer. These prior values are combined with $\mu_{e,i}$ and $\sigma_{e,i}$ through Eq. \ref{eqFac2} to obtain the posterior estimates $\mu_{d,i}$ and $\sigma_{d,i}$ from which $z_{i}$ is sampled. Finally, the value of $z_1$ is passed to a 3D convolutional decoder which aims to reconstruct the input segmentations $x$ through a series of upsampling and convolutional layers. After every convolutional and upsampling layer used in the architecture \emph{ReLu} was applied as non-linearity, except at the output of the network where \emph{sigmoid} was applied. All the network weights were randomly initialised from a zero-mean Gaussian distribution ($\sigma=0.02$).
 
 The training loss function of the LVAE+MLP network is composed of three contributions: 1) two LV segmentation reconstruction accuracy terms at ED and ES as the overlap (Dice score) between the input segmentation $x$ and its reconstruction $x'$; 2) $L$ KL divergence terms, penalising discrepancies between the learned prior and posterior distributions at each level and 3) a binary classification cross entropy (CE) term for the classification of healthy versus HCM segmentations. All the $KL_i$ divergence terms except the one of the highest level ($i=3$) were evaluated between the prior distribution $\mathcal{N}(\mu_{p,i},\sigma_{p,i}^2)$ and their posterior distribution $\mathcal{N}(\mu_{d,i},\sigma_{d,i}^2)$, while for the highest level the prior distribution was assumed to be a standard Gaussian $\mathcal{N}(0,1)$. The total loss function is 
\begin{equation}
\mathcal{L} = DSC_{ED} + DSC_{ES} + \gamma \; \Big[ \sum_{i=0}^{L} \; \alpha_i \; KL_i + \beta \; CE \Big]
\label{tottal}
\end{equation}
 and depends on $\alpha_i$, which weights the KL terms, on $\beta$, which weights the classification loss, and on $\gamma$, which is set to increase from 0 to 1 at the beginning of the training. This increase of $\gamma$ is called deterministic warm-up and it has been commonly found useful in practice to converge to better local minima \cite{sonderby2016ladder}. The weighting of the KL terms and the use of the Dice Score as a reconstruction metric lead to a different lower bound than standard VAE and LVAE. In the literature, it has been shown that the use of variants of the VAE lower-bound tend to favor better empirical results in various problems \cite{rainforth2018tighter}. In this work, we adopted Dice score as reconstruction metric since it was successfully used in our previous work \cite{biffi2018learning} and in related work \cite{cerrolaza20183d} to achieve better reconstruction results on 3D anatomical segmentations. 
 
 At testing, a pair of ED and ES LV segmentations are reconstructed by starting from $z_3=\mu_{d,3}$ and by assigning to $z_2$ and $z_1$ the values $\mu_{d,2}$ and $\mu_{d,1}$ computed from $z_3=\mu_{d,3}$ and $z_2=\mu_{d,2}$, i.e. no sampling is performed from the posterior distribution at each level. To interpret the anatomical information encoded by the highest latent space, at each level $i \ne 3$, the value of $\mu_{p,i}$ can be assigned to $z_i$ instead of $\mu_{d,i}$ and the segmentations are reconstructed as explained above. In this way, by varying the values of $z_3$, a set of segmentations at ED and ES can be directly generated for each point in $z_3$, without using the inference information provided by $\mu_{e,i}$ and $\sigma_{e,i}$. This enables the visualisation of the anatomical information encoded by the highest latent space. Finally, in order to visualise the distribution of a set of segmentations under exam in the highest latent space, the $\mu_{e,3}$ values of each segmentation can be computed through the inference network and directly plotted in a 2D space. 
 
 \begin{figure*}[h!]
\centering
\includegraphics[scale=0.36]{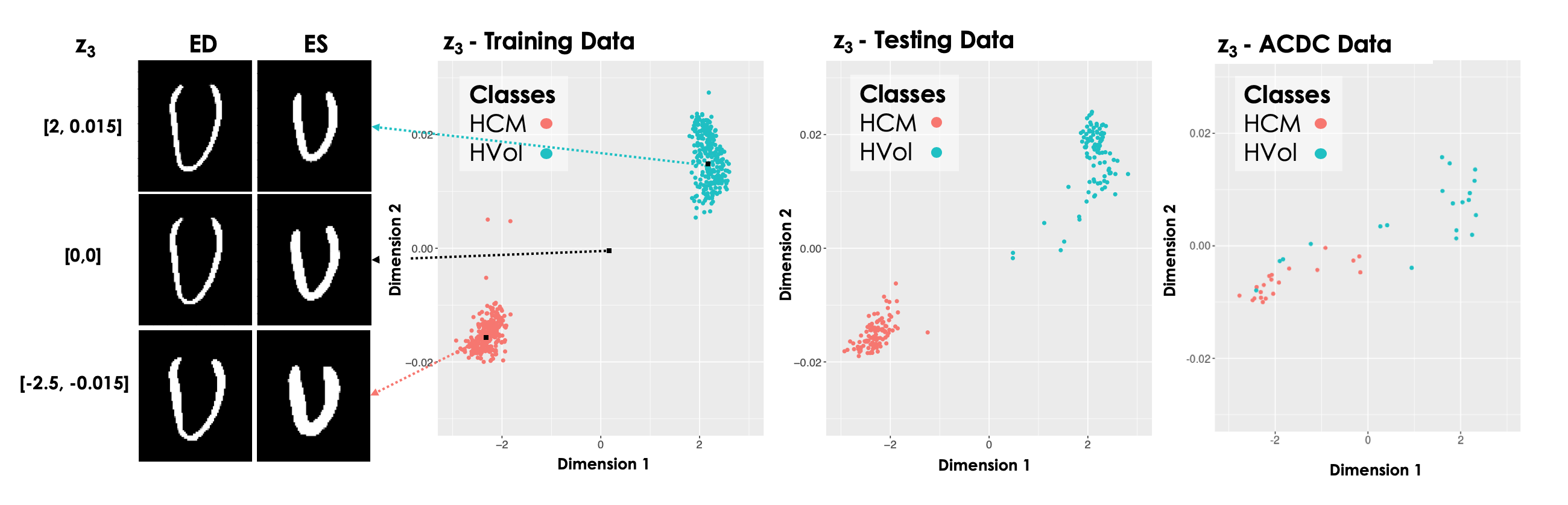}
\caption{Latent space clusters in the highest latent space ($l=3$) obtained by the proposed LVAE+MLP model on both the in-house training and testing datasets as well as on the ACDC dataset (entirely used as an additional testing dataset). Dimension 1 and 2 represent the two dimensions of $\mu_{e,3}$. On the left, long-axis sections of the reconstructed 3D segmentations at ED and ES obtained by sampling from three points in $z_3$ are shown.}
\label{latentz2}
\end{figure*}
 
\section{Results}
\subsection{Cardiac application}
\subsubsection*{Model Training}
Our in-house dataset of segmentations from healthy and HCM subjects was randomly divided into train, validation and test sets consisting of a total of 537 (276 from healthy volunteers, 261 from HCMs), 150 (75 from healthy volunteers, 75 from HCMs) and 200 (100 from healthy volunteers, 100 from HCMs) segmentations. We adopted a 3-level LVAE+MLP model (Fig. \ref{fig:LVAEarchi}) since adding more levels did neither improve the reconstruction accuracy nor the classification accuracy in the clinical application under exam. The model was trained on a NVIDIA Tesla K80 GPU using Adam optimiser with learning rate equal to $10^{-4}$ and batch size of 16. For the first 40k iterations, data augmentation including rotations around the three standard axis with rotation angles randomly extracted from a Gaussian distribution $\mathcal{N}(0,6\degree)$ was applied in order to take into account small mis-registrations. This helped the final model to achieve higher reconstruction accuracy, as it can be seen in the tables reported in the Supplementary Data 1. In the loss function (Eq. \ref{tottal}), the KL weights were fixed to $\alpha_1=0.02$, $\alpha_2=0.001$ and $\alpha_3=0.0001$ while $\gamma$ was set to increase from 0 to 100 by steps of 0.5 every 4k iterations. The relative magnitude and ascending order of the KL weights $\alpha_i$ were chosen as they provided the best segmentation reconstruction results (i.e. higher Dice Score). In particular, our experiments showed that an ascending order of the weights improves both classification and reconstruction accuracy in contrast with models having all the weights $\alpha_i$ equal or in descending order (results are shown in  Supplementary Data 2). This suggests the higher levels of a LVAE might be more difficult to train, and that a lower KL regularization term helps the training. The model produced similar results when varying these parameters within one order of magnitude, while a further increase in value reduced reconstruction accuracy and a further decrease resulted in model overfitting. The classification loss function weight $\beta$ was instead set to 0.005: we observed that a higher value would have still produced a good model, but at the price of a more unstable training at the early stages. With regards to the number of layers and nodes adopted in the MLP, we have noticed that in general adopting a single fully connected layer poses a strong constraint on the latent space distribution, while using more than two causes overfitting. The increase of the classification and KL divergence weights during training through the $\gamma$ parameter, known as deterministic warm-up \cite{sonderby2016ladder}, proved to be crucial to construct an expressive generative model (see in-depth analysis in Supplementary Data 1). After 220k iterations the training procedure was stopped as the increase of the KL divergence started to interfere with the decrease of the reconstruction and classification losses. In particular, this is due to the fact that in the highest latent space the KL divergence term tries to cluster all the data together, while the classification loss tries to separate the clusters. Hence the relative weight of $\beta$ and $\alpha_3$ needs to be tuned in order to obtain a good equilibrium. 

\begin{figure*}[h!]
\centering
\includegraphics[scale=0.45]{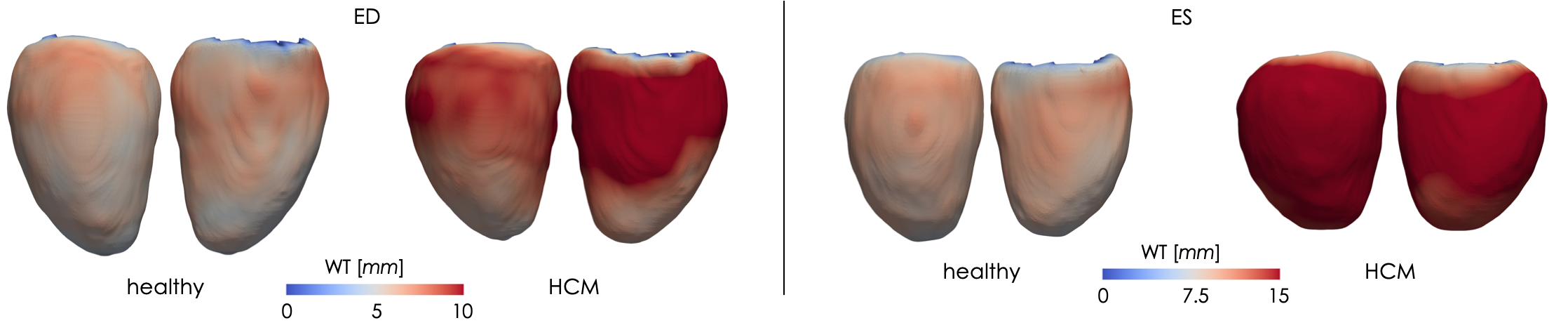}
\caption{Average healthy and HCM shapes at ED and ES sampled from the two clusters in the highest latent space of proposed LVAE+MLP model. The colormap encodes the vertex-wise wall thickness (WT), measured in mm.}
\label{meshes}
\end{figure*}

\begin{figure}[h!]
\centering
\includegraphics[width=1\columnwidth]{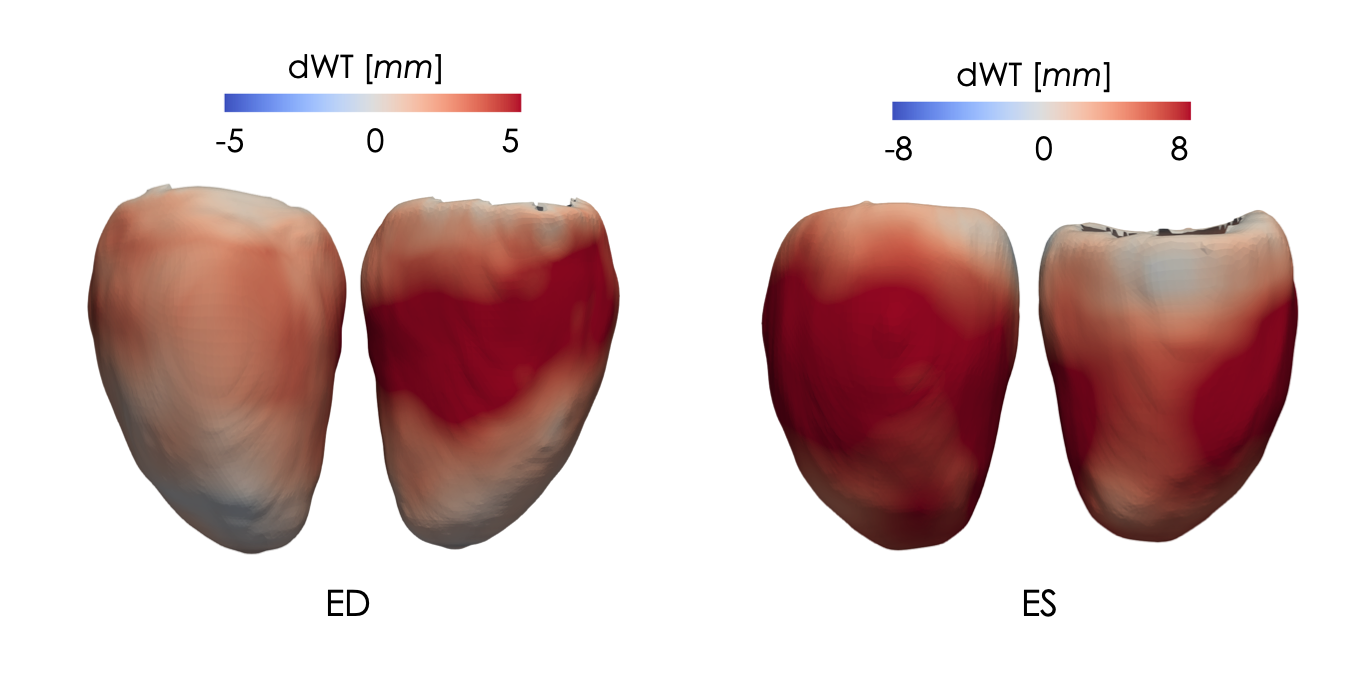}
\caption{Point-wise difference in wall thickness (dWT) at ED and ES between the healthy and the HCM average shapes of Fig. 4. Left - lateral wall; Right - septal wall.}
\label{fig:dwt}
\end{figure}

\begin{table}[!t]
\begin{minipage}[b]{1.0\linewidth}
\centering
\resizebox{9cm}{!}{
\begin{tabular}{r c c c c}
\toprule
\multicolumn{5}{c}{\textbf{VAE+MLP vs LVAE+MLP Reconstruction Accuracy}}\\
\midrule
& $DSC_{ED}$ & $DSC_{ES}$ & $H_{ED} [mm]$ & $H_{ES} [mm]$\\
\midrule
VAE+MLP train & $0.81\pm0.04$ & $0.85\pm0.04$ & $6.30\pm1.25$ & $5.96\pm1.20$\\
LVAE+MLP train & $0.85\pm0.04$ & $0.88\pm0.03$ & $5.70\pm1.12$ & $5.58\pm1.00$\\
\midrule
VAE+MLP test & $0.78\pm0.04$ & $0.83\pm0.04$ & $6.98\pm1.65$ & $6.75\pm1.61$\\
LVAE+MLP test & $0.81\pm0.04$ & $0.85\pm0.04$ & $6.54\pm1.62$ & $6.40\pm1.56$\\
\bottomrule
\end{tabular}
}
\caption{Cardiac. Dice score (DSC) and average 2D slice-by-slice Hausdorff distance (H) at ED and ES and their standard deviation for the proposed LVAE+MLP model and for the VAE+MLP model proposed in \cite{biffi2018learning} on training and testing sets.}
\label{tab:reg_res}
\end{minipage}
\end{table}

\subsubsection*{Classification and Reconstruction Results} All the 200 subjects in our testing dataset were correctly classified (100\% sensitivity and specificity) by the trained prediction network. The same model also correctly classified 36 out of the 40 ACDC MICCAI 2017 segmentations (100\% sensitivity and 80\% specificity) without the need of any re-training procedure. Of note, 3 of the 4 misclassified segmentations suffered from a lack of coverage of the LV apex, which might be the cause for the error. The results obtained for the exemplar clinical application are shown in Fig. \ref{latentz2}, where two separated clusters of segmentations have been discovered both on the training and on the testing data. An analogous result was obtained in our previous work \cite{biffi2018learning}: however, the previous version of the model required an additional dimensionality reduction step to visualise in 2D the obtained latent space of segmentations, while in the new proposed framework the highest latent space is 2D by design. Moreover, the new model achieved higher reconstruction accuracy than the previous model, as shown in Table \ref{tab:reg_res}, suggesting that a better generative model of shapes was learned. In particular, the table shows the reconstruction accuracy in terms of 3D Dice score and average 2D slice-by-slice Hausdorff distance between the 3D original and reconstructed segmentations on the testing and training datasets obtained by the proposed LVAE+MLP model and our previous VAE-based model (VAE+MLP) \cite{biffi2018learning}. The VAE+MLP model was constructed with the same 3D convolutional encoder and decoder networks of the LVAE+MLP model and with a single latent space composed of 98 latent variables, which corresponds to the total number of latent variables adopted in the LVAE+MLP model (three levels of 64, 32 and 2 latent variables, respectively). As it can be noticed in the table, the obtained Dice score results at ES are better than at ED for all the models, while the Hausdorff results seem to follow instead an opposite trend. This is probably due to the fact that since the LV is more compact at ES, the Dice score might not be sensitive to small misalignment of the reconstructed shape, which are instead captured by the Hausdorff distance.

\begin{figure}[!t]
\centering
\includegraphics[width=1\columnwidth]{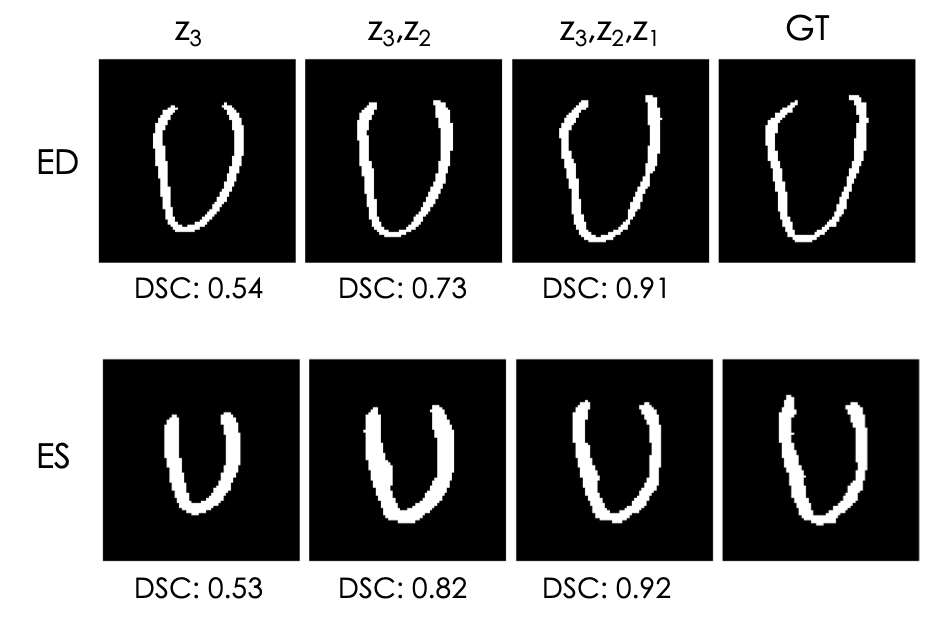}
\caption{Long-axis section of reconstructed segmentations at ED and ES by the LVAE+MLP model, using only $z_3$ information (first column) or also using the posterior information of the other latent spaces $(z_2, z_1)$ . Last column: ground-truth (GT) segmentation. DSC = Dice Score between the segmentation at that column and the GT.}
\label{fig:oneRec}
\end{figure}

\subsubsection*{Visualisation of the latent spaces} 
\begin{figure}[!t]
\centering
\includegraphics[width=0.95\columnwidth]{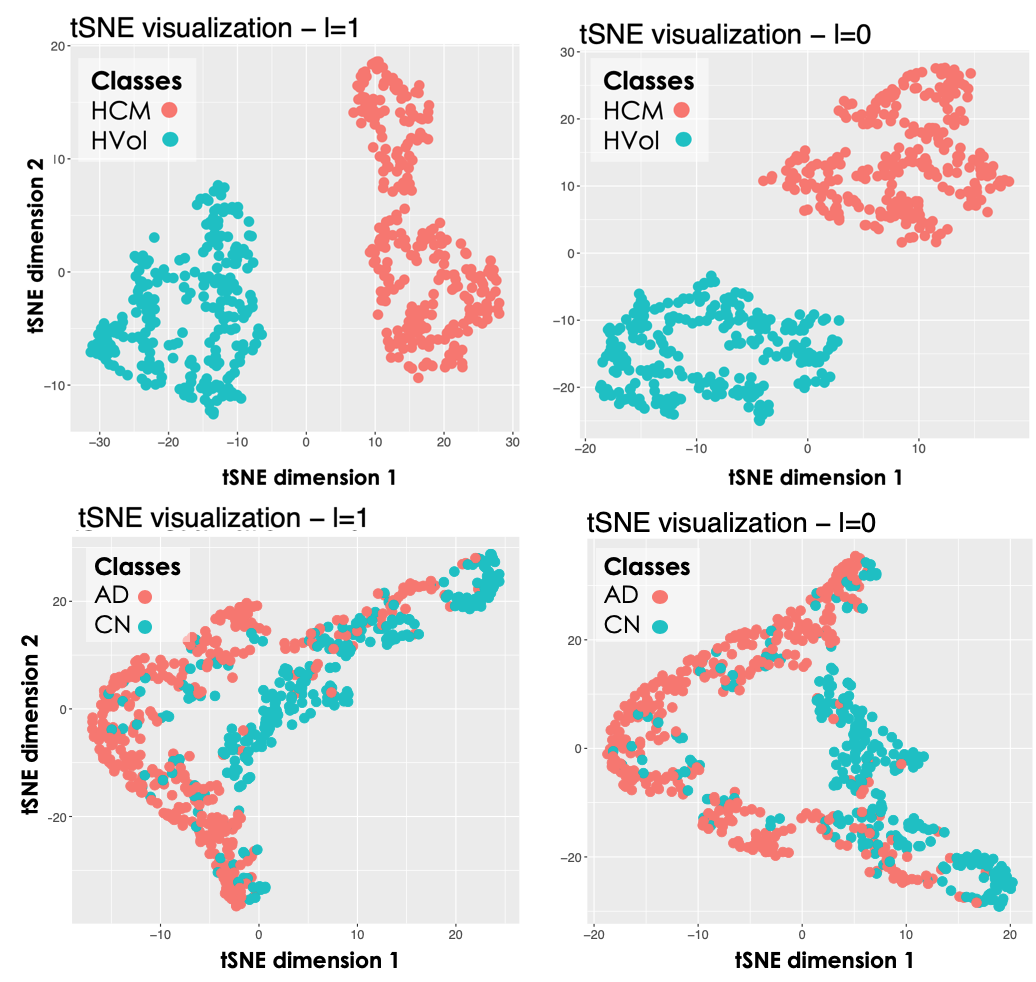}
\caption{tSNE visualisation of the latent spaces $z_2$ and $z_1$. Top: cardiac application. Bottom: brain application.}
\label{l1l0}
\end{figure}

Thanks to the properties of the proposed model, the anatomical information encoded by each latent space can be directly visualised, especially the information embedded in the highest level ($i=3$), which encodes the most discriminative features for the classification of healthy and HCMs 3D LV segmentations. For the exemplar application under investigation, little intra-cluster variability between the shapes generated from the latent space $z_3$ was obtained, while much larger inter-cluster variability between the generated shapes was obtained. This can be seen on the left-side of Fig. \ref{latentz2} where we report a long-axis section of the 3D reconstructed segmentations at ED and ES at three points of the latent space $z_3$ (for a grid visualisation of the shapes encoded by this latent space please refer to Supplementary Data 3). Moreover, in Fig. \ref{meshes} we show the obtained mean average shape for each cluster, represented as a triangular mesh with point-wise wall thickness (WT) values at vertex. This was obtained by sampling $N=1000$ segmentations from each cluster in $z_3$ after estimating its probability density via kernel density estimation. Then, the obtained segmentations for each cluster were averaged to extract the corresponding average segmentation. Finally, a non-rigid transformation between the obtained average segmentation and a 3D high-resolution LV segmentation from the UK Digital Heart project\footnote{https://digital-heart.org/} was computed, and the inverse of this transformation was applied to the corresponding 3D high-resolution LV segmentation to warp it to the cluster specific average segmentation. At each of the mesh vertices, WT was then computed as the perpendicular distance between the endocardial and epicardial wall. The results are presented in Fig. \ref{meshes}, where it can be noticed that the average HCM shape has higher WT than the corresponding healthy shape and it has a slightly reduced size. Fig. \ref{fig:dwt} instead reports the point-wise difference in WT between the HCM and the healthy shape, and it can be noticed that the most discriminative anatomical feature to classify an HCM shape consists in an increased WT in the septum, which is in agreement with the clinical literature \cite{desai2011imaging}. Fig. \ref{fig:oneRec} shows a long-axis section of the reconstructed segmentations at ED and ES from the LVAE+MLP model when only $z_3$ posterior information is used (first column) or when also the posterior information in the other levels ($z_2, z_1)$ is exploited: the latent spaces $z_2$ and $z_1$ evidently encode different anatomical features that help to refine  the structural information provided by $z_3$. Results for more subjects are reported in Supplementary Data 4. Finally, we applied the dimensionality reduction technique tSNE \cite{maaten2008visualizing} to visualise in two dimensions the distributions of $z_1$ and $z_2$ latent spaces and we have found that the latent representations of the two classes of shapes are clustered also at both these levels (plots shown in Fig. \ref{l1l0}). A possible explanation relies on the fact that the generative process is a conditional: if the data is clustered at the top of the hierarchy, it may be easier for the network to keep the clusters also in the subsequent levels. 

\subsection{Brain application}
\subsubsection*{Model Training} As an additional benchmark test, the LVAE+MLP model proposed in this work was also tested for the classification of healthy controls (HC) and patients with AD by using only 3D segmentations of the left and right hippocampus. Data was randomly assigned to train, validation and testing sets consisting of a total of 562 (322 HC, 240 AD), 64 (32 HC, 32 AD) and 100 (50 HC, 50 AD) segmentations respectively. A three level LVAE+MLP model was also adopted for this application (scheme in Supplementary Materials 5), since adding more levels did not improve classification or reconstruction accuracy. In the loss function (Eq. \ref{tottal}), the KL weights were fixed to $\alpha_1=0.03$, $\alpha_2=0.003$ and $\alpha_3=0.0003$, $\gamma$ was set to increase from 0 to 100 by steps of 0.5 every 4k iterations and $\beta$ was instead set to 0.005. The same augmentation strategy and the rationale for the selection of the hyperparameters in the previous experiment were adopted.  The model training was stopped after 200k iterations.

\begin{figure*}[h!]
\centering
\includegraphics[scale=0.41]{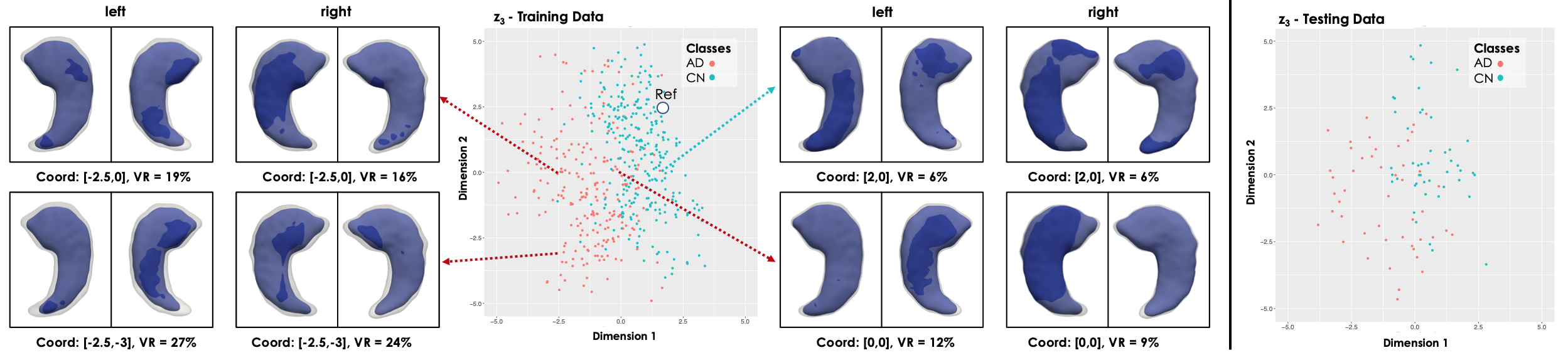}
\caption{Latent space clusters in the highest latent space ($l=3$) obtained by the proposed LVAE+MLP model on the brain dataset. Left and right hippocampus shapes (in blue) at four points in the latent space have been reconstructed and showed together with a reference shape (in grey and opaque) sampled from the healthy control shapes (Ref, Coord: [2,2]). The first image is a view from the top, second image a view from the bottom.}
\label{latentzADNI}
\end{figure*}

\begin{figure}[h!]
\centering
\includegraphics[width=0.9\columnwidth]{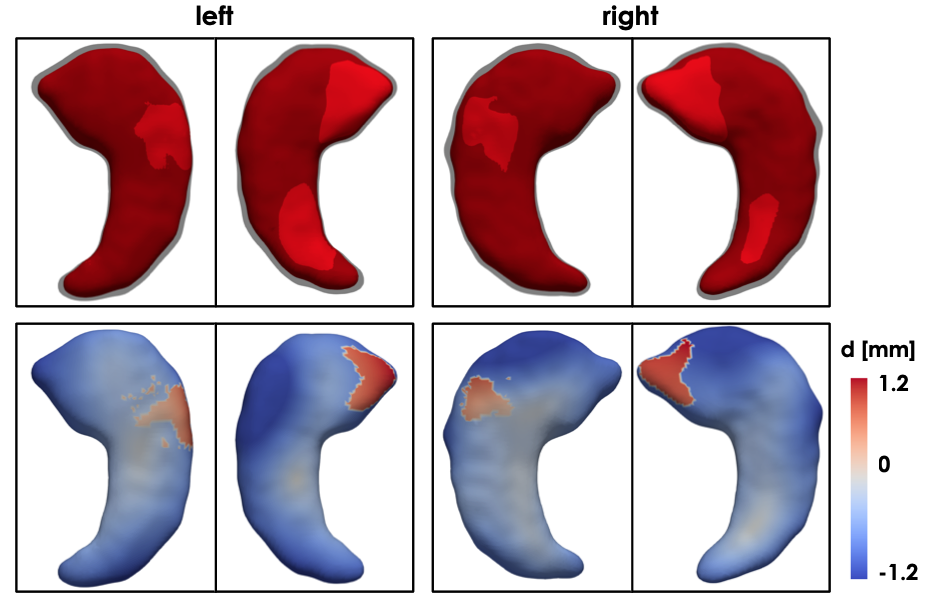}
\caption{First row: Average healthy (in grey and opaque) and AD (in red) left and right hippocampus shapes sampled from the two clusters in the highest latent space of proposed LVAE+MLP model. Second row: vertex-by-vertex L2 distance between the two mean shapes.}
\label{fig:mean_shapes_ADNI}
\end{figure}

\begin{table}[!t]
\begin{minipage}[b]{1.0\linewidth}
\centering
\resizebox{9cm}{!}{
\begin{tabular}{r c c c c}
\toprule
\multicolumn{5}{c}{\textbf{VAE+MLP vs LVAE+MLP Reconstruction Accuracy}}\\
\midrule
& $DSC_{l}$ & $DSC_{r}$ & $H_{r} [mm]$ & $H_{l} [mm]$\\
\midrule
VAE+MLP train & $0.81\pm 0.05$ & $0.80\pm0.05$ & $3.35\pm0.67$ & $3.28\pm0.69$\\
LVAE+MLP train & $0.85\pm 0.04$ & $0.85\pm 0.03$ & $3.05\pm0.69$ & $2.96\pm0.66$\\
\midrule
VAE+MLP test & $0.79\pm0.05$ & $0.79\pm0.05$ & $3.51\pm0.64$ & $3.49\pm0.67$\\
LVAE+MLP test & $0.82\pm0.03$ & $0.82\pm0.03$ & $3.31\pm0.68$ & $3.23\pm0.65$\\
\bottomrule
\end{tabular}
}
\caption{Brain. Dice score (DSC) and average 2D slice-by-slice Hausdorff distance (H) for the left (l) and right (r) hippocampus and their standard deviation for the proposed LVAE+MLP model and for the VAE+MLP model proposed in \cite{biffi2018learning} on training and testing sets.}
\label{tab:reg_res2}
\end{minipage}
\end{table}

\subsubsection*{Classification and reconstruction results} 
84 out of 100 subjects were correctly classified by the training prediction network (78\% sensitivity, 90\% specificity). A VAE+MLP model with the same 3D convolutional encoder and decoder networks of the LVAE+MLP model, but with a single latent space of dimension 66 (equal to the total number of latent variables adopted in the LVAE+MLP model) was also trained. This model classified 81 out of 100 subjects correctly (74\% sensitivity, 88\% specificity) on the same training, testing and validation splits of the previous model. On the same dataset, an accuracy of 78\% (75\% sensitivity, specificity 80\%) for the same classification task was obtained by using left and right hippocampus volume segmentations \cite{ledig2018structural}. Compared to the VAE+MLP model, the LVAE+MLP model achieves higher reconstruction accuracy in terms of 3D Dice score and 2D slice-by-slice Hausdorff distance between the original segmentations and the reconstructed ones, these results are reported in Table \ref{tab:reg_res2}. 

\subsubsection*{Visualisation of the latent spaces}
Fig. \ref{latentzADNI} shows the distribution of the training and testing 3D hippocampus segmentations in the highest ($i=3$) latent space for the trained LVAE+MLP model. It can be noticed how the healthy and pathological shapes are not as separated as in the previous application due to the more challenging nature of the new task. However, two clear clusters of healthy and AD shapes can still be identified. Fig. \ref{latentzADNI} also shows the left and right hippocampus segmentations obtained by sampling from four distinct points of this latent space, which are showed together with a reference healthy shape sampled from a point in the healthy cluster (marked as Ref). For each reconstructed segmentation, the rate of hippocampal volume change (VR) with respect to the reference healthy shape was computed ($VR = \left| \frac{V-V_{ref}}{V_{ref}}\ \right| \times 100 $). From the figure, it can be noticed how the AD shapes are characterized by decreased hippocampal volume, reduction that slighty but consistently affects more the left than the right hippocampus, in agreement with the previous findins on this data~\cite{ledig2018structural}. Moreover, a pattern in regional changes in volume can be identified: AD cases closer to the reference healthy shape show atrophy predominantly (if not only) in the tail of the hippocampus, while cases further away from the healthy class and deeper into the AD group show an atrophy pattern more spread throughout the whole hippocampal shape. In Fig. \ref{fig:mean_shapes_ADNI}, we show the obtained average left and right hippocampus shapes from the healthy and AD distribution represented as triangular meshes. These meshes were obtained by sampling $N=1000$ segmentations from the healthy and AD distributions in $z_3$ after estimating their probability density via kernel density estimation. Then, the obtained segmentations for each cluster were averaged to extract the corresponding cluster mean segmentation. Finally, the 3D template hippocampus segmentation was non-ridigly registered to each obtained cluster specific average segmentation and the estimated transformation was applied to the corresponding high-resolution mesh. In the first row of Fig. \ref{fig:mean_shapes_ADNI}, it can be noticed how the reconstructed template AD segmentation (red) which is shown together with the HC segmentation (grey and opaque) is more atrophied and it is characterized by a bending of the head of both the left and right hippocampus. The second row displays the vertex-by-vertex L2 distance between the two mean shapes demonstrating a more pronounced regional atrophy in the hippocampal head consistent with the CA1 and subiculum regional athrophy already reported in the literature \cite{shen2012detecting}, \cite{mueller2010hippocampal}. The right hippocampus is characterized by a 13.5\% decrease in volume between the healthy shape and the AD shape, while the decrease in volume for the left hippocampus is 14.6\%. The volume ratio between the AD right and left hippocampus is 3.6\% and 2.5\% in the healthy mean shape. Finally, the plots resulting from the application of tSNE dimensionality reduction technique to the $z_1$ and $z_2$ training data values are shown at the bottom of Fig. \ref{l1l0}.

\section{Discussion}
In this work, we presented a data-driven framework which learns to model a population of 3D anatomical segmentations through a hierarchy of conditional latent variables, encoding at the highest level of the hierarchy the most discriminative features to differentiate distinct clinical conditions. This is achieved by implementing and extending for the first time the LVAE framework to a real-world medical imaging application. In particular, by performing a classification task in the highest level of the LVAE hierarchy of latent variables, we can force this latent space to encode the most discriminative features for a clinical task under exam, while the other levels encode the other factors of anatomical variation needed to model the manifold of segmentations under analysis. Being a generative model, this framework provides the advantage of enabling the visualisation and quantification of the remodeling effect encoded by each latent space in the original segmentation space. Hence, the anatomical differences used by the classifier to distinguish different conditions can be easily visualised and quantified by sampling from the highest level posterior distribution computed from a given database of shapes. Moreover, by designing this latent space to be two or three dimensional, no additional offline dimensionality reduction technique is required to visually assess the distribution of these shapes in the latent space. As a consequence, this method not only provides a deep learning classifier that uses a task-specific latent space in the discrimination of different clinical conditions, but more importantly enables the visualisation of the anatomical features encoded by this latent space, making the classification task transparent.

With the aim of assisting the clinicians in quantifying the morphological changes related to disease, we have applied the proposed framework for the automatic classification of heart and brain pathologies against healthy controls. In the reported cardiac application, the learned features achieved high accuracy in the discrimination of healthy subjects from HCM patients on our unseen testing dataset and on a second external testing dataset from the ACDC MICCAI 17 challenge. On the more challenging task of classification of healthy versus AD hippocampi, the model achieved better classification accuracy than using volumetric indices \cite{ledig2018structural} and our previous method. Moreover, the visualisation of the features encoded in the highest level of the adopted LVAE+MLP model showed how the proposed model is able to provide the clinicians with a 3D visualisation of the most discriminative anatomical changes for the task under study, making the data-driven assessment of regional and asymmetric remodelling patterns characterizing a given clinical condition possible. 

On both applications, we have also showed that the proposed LVAE+MLP model allows the construction of a better generative model in comparison to a VAE-based model with a single latent space  \cite{biffi2018learning}. To the best of our knowledge, this result confirms for the first time that hierarchical latent spaces provide a more accurate generative model on a real clinical dataset. Moreover, this work also gives insights on the functioning of these models on 3D anatomical segmentations, including how the different levels of latent variables encode different anatomical features, and how to optimally train this class of models for the reconstruction of these 3D anatomical segmentations.

While this work showed the potentialities of the proposed method on two common classification tasks, this method is domain-agnostic and could be applied to other classification problems where pathological remodelling is a predictor of a disease class label. However, further work is needed to explore the full potential of this approach, for instance in order to visualize the pathological remodeling of different disease subgroups characterized by different clinical endpoints. Of note, we expect that on very difficult tasks one or two more dimensions in the highest latent space might be needed, although further increasing the dimensionality will go against the rationale of the proposed approach. 
In fact, our aim is to encode the most discriminative anatomical information for the classification task under exam in the highest latent space, while the other latent spaces are intentionally left to model the remaining factors of variation. Interestingly, Fig. \ref{l1l0} shows that the shapes are clustered also in the other latent spaces, probably encoding additional variability of the disease groups not useful for the specific classification task.  By specializing the classification task to more categories, we expect some information currently encoded in the other latent spaces to be moved and encoded in the highest one. For instance, studying multiple disease subgroups would enable a finer representation of the spectrum of remodeling patterns against which patients can be compared. Presently this was not achievable as the model has been optimized to discriminate only between healthy and diseased subjects, although a step in this direction was taken in Fig. \ref{latentzADNI}, showing how different latent space points map to different hippocampal volume measures. 

In comparison with the previous (generative) model \cite{biffi2018learning} and Bello \emph{et al.} \cite{bello2019deep} model, the proposed method requires tuning of a few additional hyperparameters, i.e. the number of adopted levels in the ladder and their weights importance in the model loss function. On the other hand, our approach is fully data-driven and it spares the need for further downstream dimensionality reduction and latent space navigation techniques, which would themselves require separate optimization and human intervention, potentially adding bias to the analysis. The proposed method also enables the derivation of population-based inferences (Fig. \ref{meshes}, \ref{fig:dwt} and \ref{fig:mean_shapes_ADNI}), which could neither have been obtained from our previous model (due to the subject-specific nature of the latent-space navigation), nor from the one of Bello \emph{et al.} (due to the non-generative nature of the model).

Another limitation shared by our previous and current approach is the fact that the input segmentations need to be rigidly registered to train the model. Future work should consider how to extend the proposed method to unregistered shapes, for example with the introduction of spatial transformer modules inside the architecture. In this work, as the output of the model is binary, Dice score was adopted as reconstruction metric. However, other alternatives exist, for example by modeling the model output with a Bernoulli distribution \cite{doersch2016tutorial}, and they will be investigated in future work. Finally, the prior distribution adopted in the highest latent space is a standard Gaussian distribution $\mathcal{N}(0,1)$: future work could consider alternative prior distributions which could further favour the clustering of shapes. Even more interestingly, the interpretability and visualisation properties of the proposed method indicate that it could constitute an interesting tool for unsupervised clustering of shapes, for example by learning in the highest level discrete random variables.  

\section{Conclusions}
In recent years, the medical image analysis field has witnessed a marked increase both in the construction of large-scale population-based imaging databases and in the development of automated segmentation frameworks. As a consequence, the need for novel approaches to process and extract clinically relevant information from the collected data has greatly increased. In this work, we proposed a method for data-driven shape analysis which enables the classification of different groups of clinical conditions through a very low-dimensional set of task-specific features. Moreover, this framework naturally enables the quantification and visualisation of the anatomical effects encoded by these features in the original space of the segmentations, making the classification task transparent. As a consequence, we believe that this method will be useful for the study of both normal anatomy and pathology in large-scale studies of volumetric imaging.

\bibliographystyle{IEEEtran}

\bibliography{bibliography}

\onecolumn

\section*{Supplementary Data 1 - Model Training - Effect of Deterministic Warm-Up (DWU) and Data Augmentation (DA)}

\begin{table}[h!]
\centering
\begin{tabular}{c c c c c c}
\hline
\multicolumn{6}{c}{\textbf{Effect of DA and DWU}}\\
\hline
\multicolumn{6}{c}{Training}\\
\hline
& $DSC_{ED}$ & $DSC_{ES}$ & $H_{ED} [mm]$ & $H_{ES} [mm]$ & ACC [\%]\\
\hline
None & $0.75\pm 0.07$ & $0.79\pm 0.05$ & $7.30\pm1.80$ & $7.08\pm1.68$ & $51.40\%$ \\
DA & $0.77\pm 0.05$ & $0.80\pm 0.05$ & $6.94\pm1.62$ & $6.86\pm1.53$ & $51.40\%$\\
DWU & $0.82\pm 0.05$ & $0.86\pm 0.04$ & $6.20\pm1.23$ & $5.93\pm1.23$  & $99\%$ \\
DA+DWU  & $0.85\pm 0.04$ & $0.88\pm 0.03$ & $5.70\pm1.12$ & $5.58\pm 1.00$ & $100\% $\\
\hline
\bottomrule
\end{tabular}
\caption{Dice score (DSC) and average 2D slice-by-slice Hausdorff distance (H) at ED and ES and their standard error for the proposed LVAE+MLP model when Deterministic Warm-Up (DWU) and Data Augmentation (DA) are applied. ACC is the classification accuracy of the different models. Results obtained on the training dataset.}
\label{tab:reg_res}
\end{table}

\begin{table}[h!]
\centering
\begin{tabular}{c c c c c c}
\hline
\multicolumn{6}{c}{\textbf{Effect of DA and DWU}}\\
\hline
\multicolumn{6}{c}{Testing}\\
\hline
& $DSC_{ED}$ & $DSC_{ES}$ & $H_{ED} [mm] $ & $H_{ES} [mm]$ & $ACC [\%]$\\
\hline
None & $0.72\pm 0.07$ & $0.76\pm 0.05$ & $8.01\pm1.99$ & $7.53\pm1.97$ &$ 51.40\% $\\
DA & $0.74\pm 0.06$ & $0.78\pm 0.05$ & $7.62\pm1.86$ & $7.31\pm1.82$ &$ 51.40\%$\\
DWU & $0.79\pm 0.05$ & $0.83\pm 0.04$ & $6.91\pm1.79$ & $6.72\pm 1.68$  & $99\%$ \\
DA+DWU  & $0.81\pm 0.04$ & $0.85\pm 0.04$ & $6.54\pm1.62$ & $6.40\pm 1.56$ & $100\%$ \\
\hline
\bottomrule
\end{tabular}
\caption{Dice score (DSC) and average 2D slice-by-slice Hausdorff distance (H) at ED and ES and they standard error of the mean for the proposed LVAE+MLP model when Deterministic Warm-Up (DWU) and Data Augmentation (DA) are applied. ACC is the classification accuracy of the different models. Results obtained on the testing dataset.}
\label{tab:reg_res}
\end{table}

\newpage

\section*{Supplementary Data 2 - Model Training - Effect of the KL weights}

\begin{table}[h!]
\centering
\begin{tabular}{c c c c c c}
\hline
\multicolumn{6}{c}{\textbf{Effect of the KL weights}}\\
\hline
\multicolumn{6}{c}{Training}\\
\hline
[$\alpha_1$,$\alpha_2$,$\alpha_3$] & $DSC_{ED}$ & $DSC_{ES}$ & $H_{ED} [mm]$ & $H_{ES} [mm]$ & $ACC [\%]$\\
\hline
$[10^{-4}, 2 \; 10^{-4}, 10^{-3}]$ & $0.79\pm 0.05 $& $0.80\pm 0.05$ & $6.93\pm1.62$ & $6.88\pm1.60$ & $100\%$\\
$[10^{-4}, 10^{-4}, 10^{-4}]$ & $0.80\pm 0.05$ & $0.83\pm 0.04$ & $6.50\pm1.41$ & $6.48\pm 1.47$ & $100\%$ \\
$[10^{-3}, 2 \; 10^{-4}, 10^{-4}]$  & $0.85\pm 0.04$ & $0.88\pm 0.03$ & $5.70\pm1.12$ & $5.58\pm 1.00$& $100\%$ \\
\hline
\bottomrule
\end{tabular}
\caption{Dice score (DSC), average 2D slice-by-slice Hausdorff distance (H) at ED and ES and they standard error of the mean together with classification accuracy (C) for the proposed LVAE+MLP model for different sets of the KL weights $\alpha_i$ in the training loss function. ACC is the classification accuracy of the different models. Results obtained on the training dataset.}
\label{tab:reg_res}
\end{table}

\begin{table}[h!]
\centering
\begin{tabular}{c c c c c c}
\hline
\multicolumn{6}{c}{\textbf{Effect of the KL weights}}\\
\hline
\multicolumn{6}{c}{Testing}\\
\hline
[$\alpha_1$,$\alpha_2$,$\alpha_3$]  & $DSC_{ED}$ & $DSC_{ES}$ & $H_{ED}$ [mm] & $H_{ES}$ [mm] & $ACC [\%]$\\
\hline
$[10^{-4}, 2 \; 10^{-4}, 10^{-3}]$ & $0.75\pm 0.06$ & $0.78\pm 0.05$ & $7.64\pm1.72$ & $7.37\pm 1.68$ & $99\%$\\
$[10^{-4}, 10^{-4}, 10^{-4}]$ & $0.78\pm 0.05$ & $0.80\pm 0.04$ & $7.01\pm1.53$ & $6.94\pm1.58$ & $100\%$ \\
$[10^{-3}, 2 \; 10^{-4}, 10^{-4}]$  & $0.81\pm0.04$ & $0.85\pm0.04$ & $6.54\pm1.62$ & $6.40\pm1.56$ & $100\%$ \\
\hline
\bottomrule
\end{tabular}
\caption{Testing data results. Dice score (DSC), average 2D slice-by-slice Hausdorff distance (H) at ED and ES and they standard error of the mean together with classification accuracy (C) for the proposed LVAE+MLP model for different sets of the KL weights $\alpha_i$ in the training loss function. ACC is the classification accuracy of the different models.  Results obtained on the testing dataset.}
\label{tab:reg_res}
\end{table}

\clearpage
\newpage
\mbox{~}

\section*{Supplementary Data 3 - Cardiac Application - Segmentations encoded by $z_3$} 

\begin{figure}[h!]
\centering
\includegraphics[width=0.8\columnwidth]{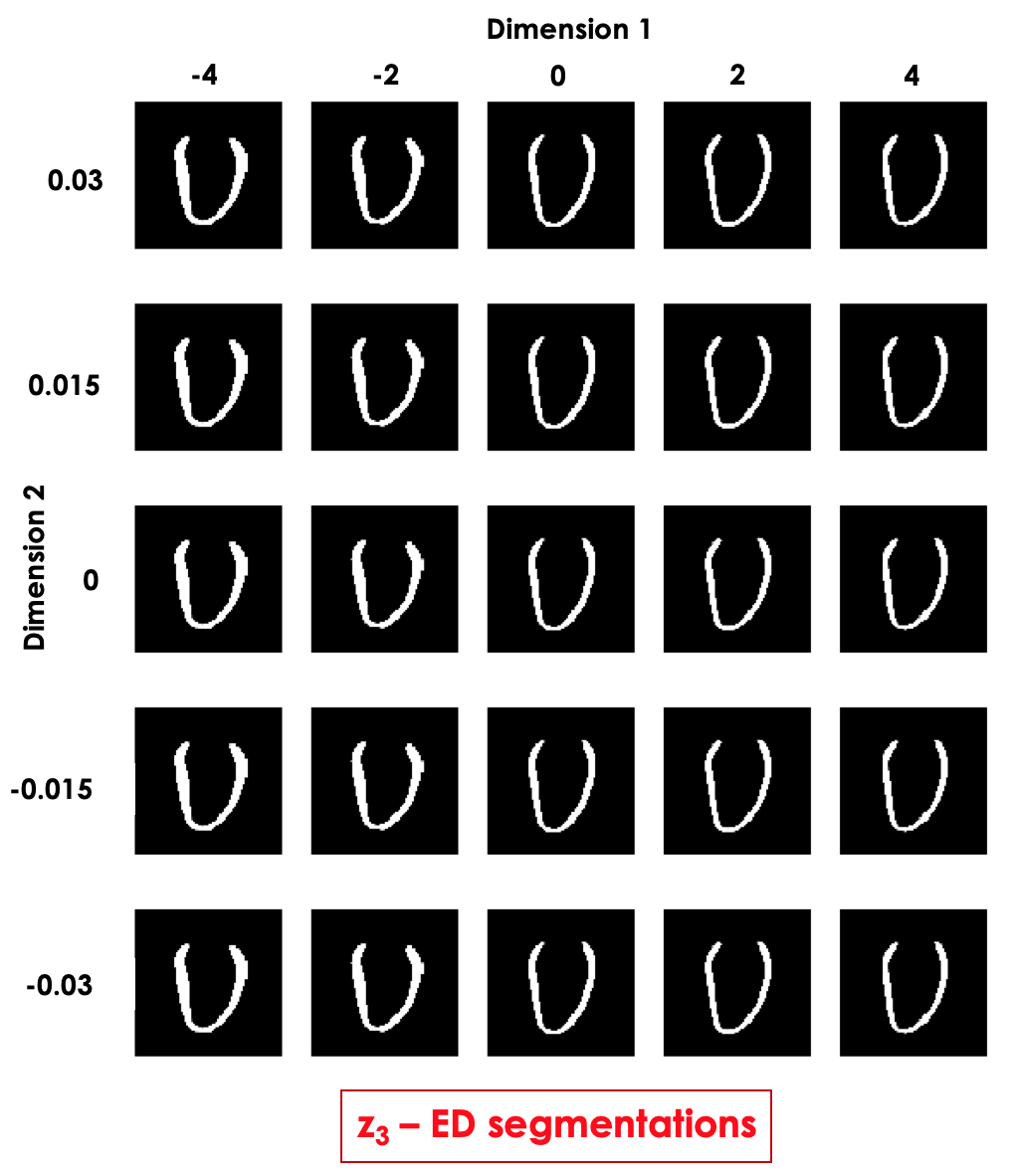}
\caption{Long-axis section of reconstructed segmentations at ED by the LVAE+MLP model by sampling from different points in $z_3$ and subsequently from the prior distribution of the latent variables $z_2$ and $z_1$.}
\label{fig:EDlatent}
\end{figure}

\begin{figure}[h!]
\centering
\includegraphics[width=0.8\columnwidth]{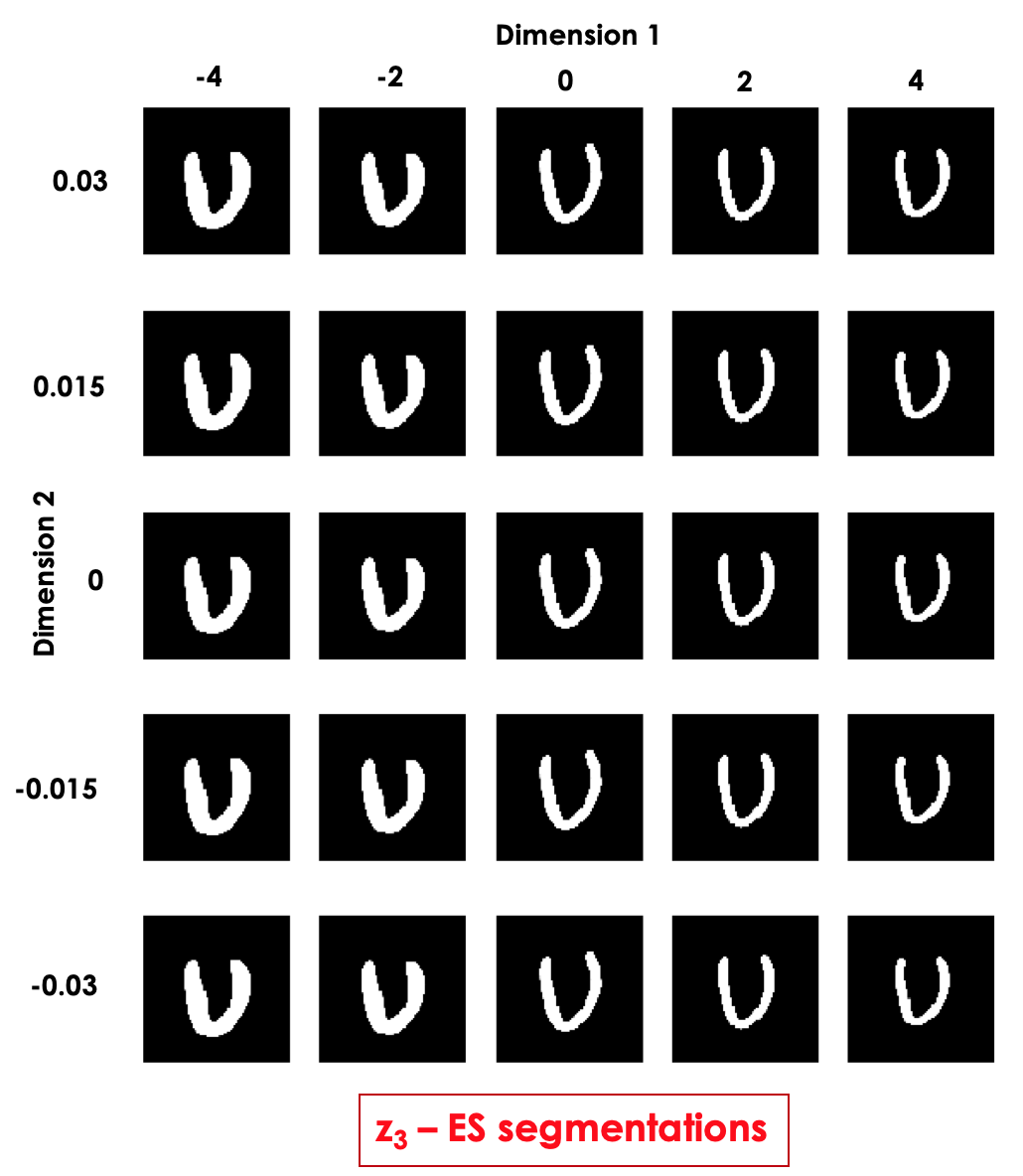}
\caption{Long-axis section of reconstructed segmentations at ES by the LVAE+MLP model by sampling from different points in $z_3$ and subsequently from the prior distribution of the latent variables $z_2$ and $z_1$.}
\label{fig:ESlatent}
\end{figure}

\clearpage

\mbox{~}

\section*{Supplementary Data 4 - Cardiac Application - Additional Reconstruction Examples} 
\begin{figure}[h!]
\centering
\includegraphics[width=0.85\columnwidth]{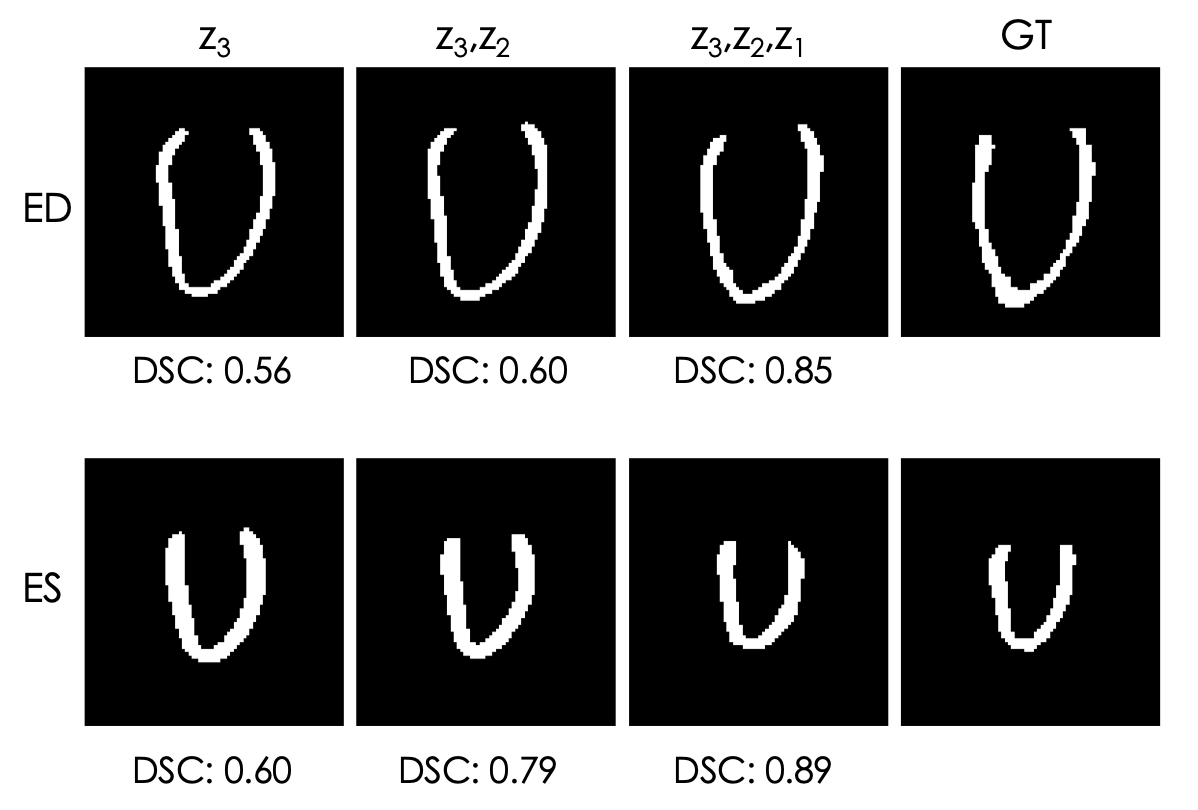}
\caption{Long-axis section of an healthy subject reconstructed segmentations at ED and ES by the LVAE+MLP model using only $z_3$ information (first column) or also using the posterior information of the other latent spaces $(z_2, z_1)$ .  Last column: ground-truth (GT) segmentation. DSC = Dice Score between the segmentation at that column and the GT.}
\label{fig:recHVOL}
\end{figure}

\newpage

\begin{figure}[h!]
\centering
\includegraphics[width=0.85\columnwidth]{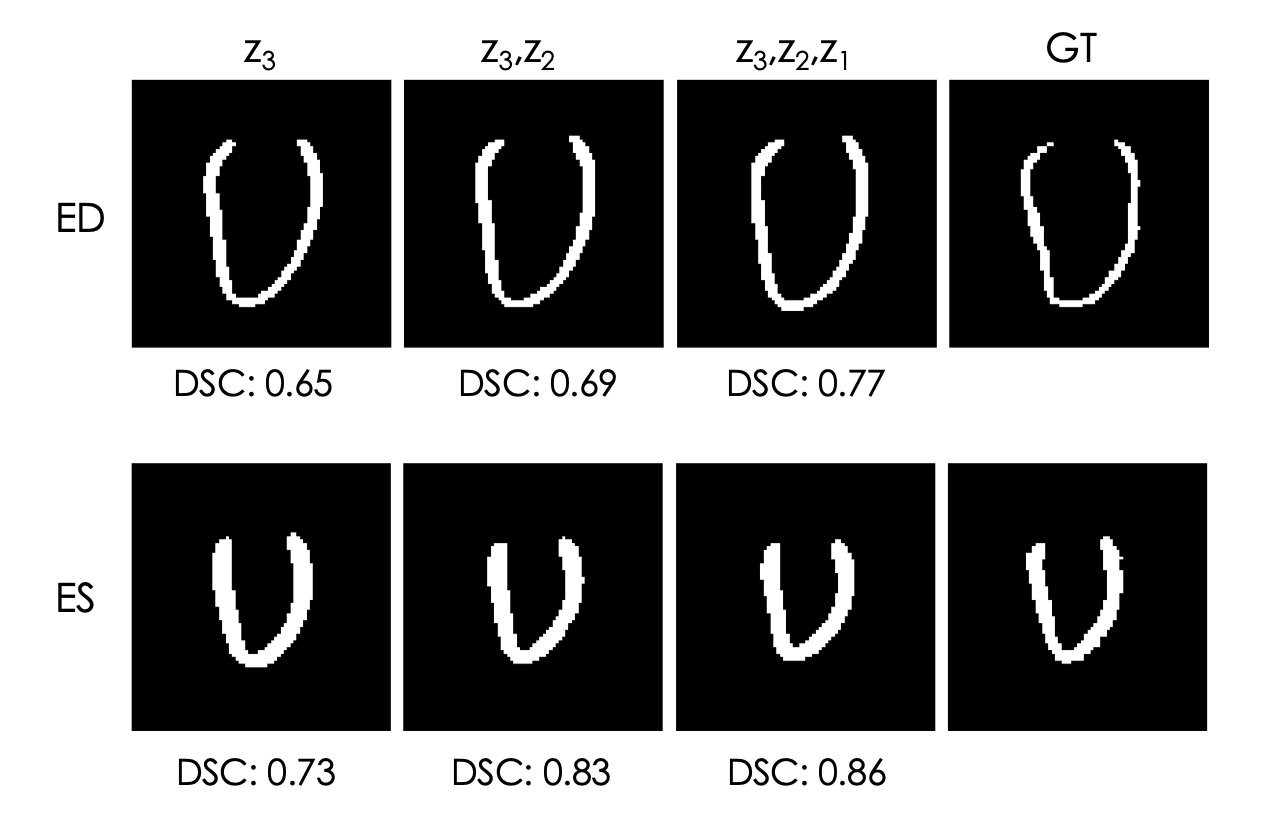}
\caption{Long-axis section of an healthy subject reconstructed segmentations at ED and ES by the LVAE+MLP model using only $z_3$ information (first column) or also using the posterior information of the other latent spaces $(z_2, z_1)$ .  Last column: ground-truth (GT) segmentation. DSC = Dice Score between the segmentation at that column and the GT.}
\label{fig:recHVOL}
\end{figure}

\newpage

\begin{figure}[h!]
\centering
\includegraphics[width=0.85\columnwidth]{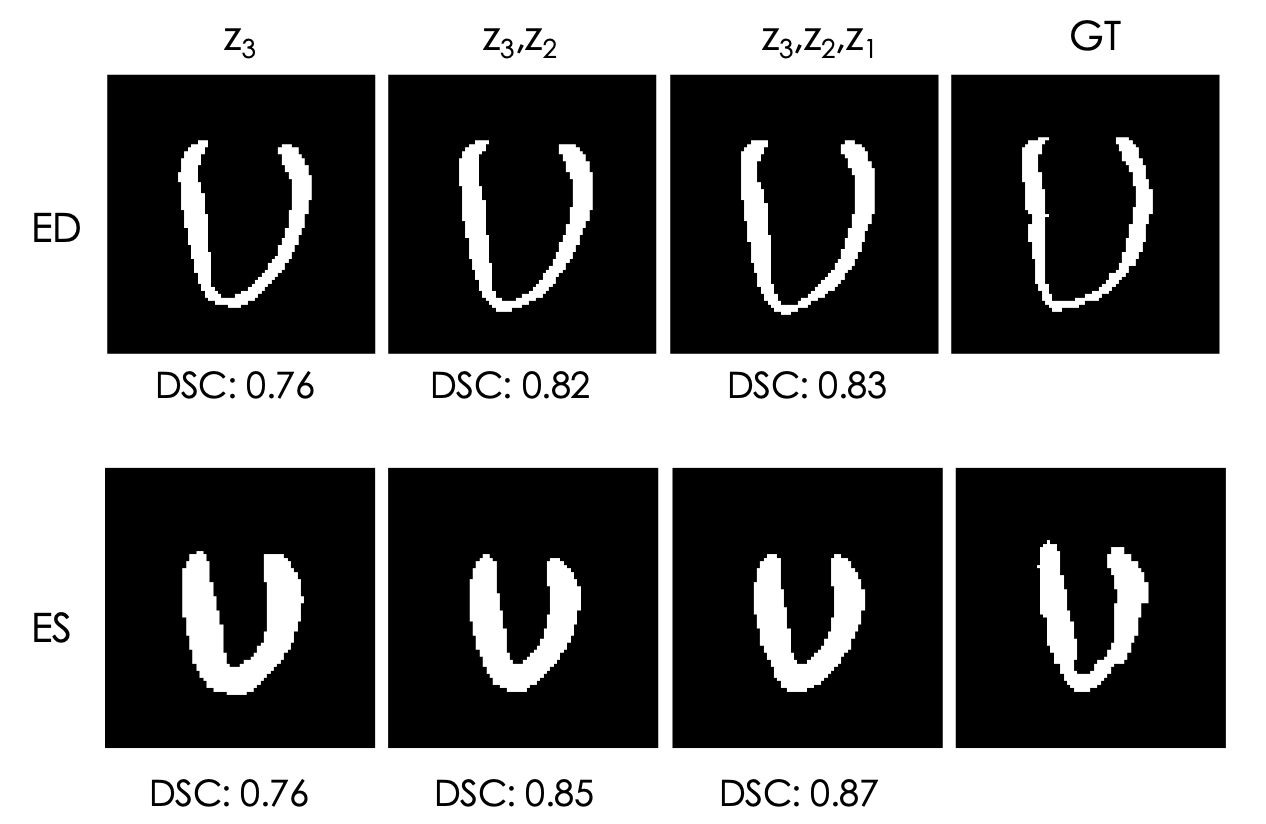}
\caption{Long-axis section of an HCM patient reconstructed segmentations at ED and ES by the LVAE+MLP model using only $z_3$ information (first column) or also using the posterior information of the other latent spaces $(z_2, z_1)$ .  Last column: ground-truth (GT) segmentation. DSC = Dice Score between the segmentation at that column and the GT.}
\label{fig:recHVOL}
\end{figure}
\newpage
\begin{figure}[h!]
\centering
\includegraphics[width=0.85\columnwidth]{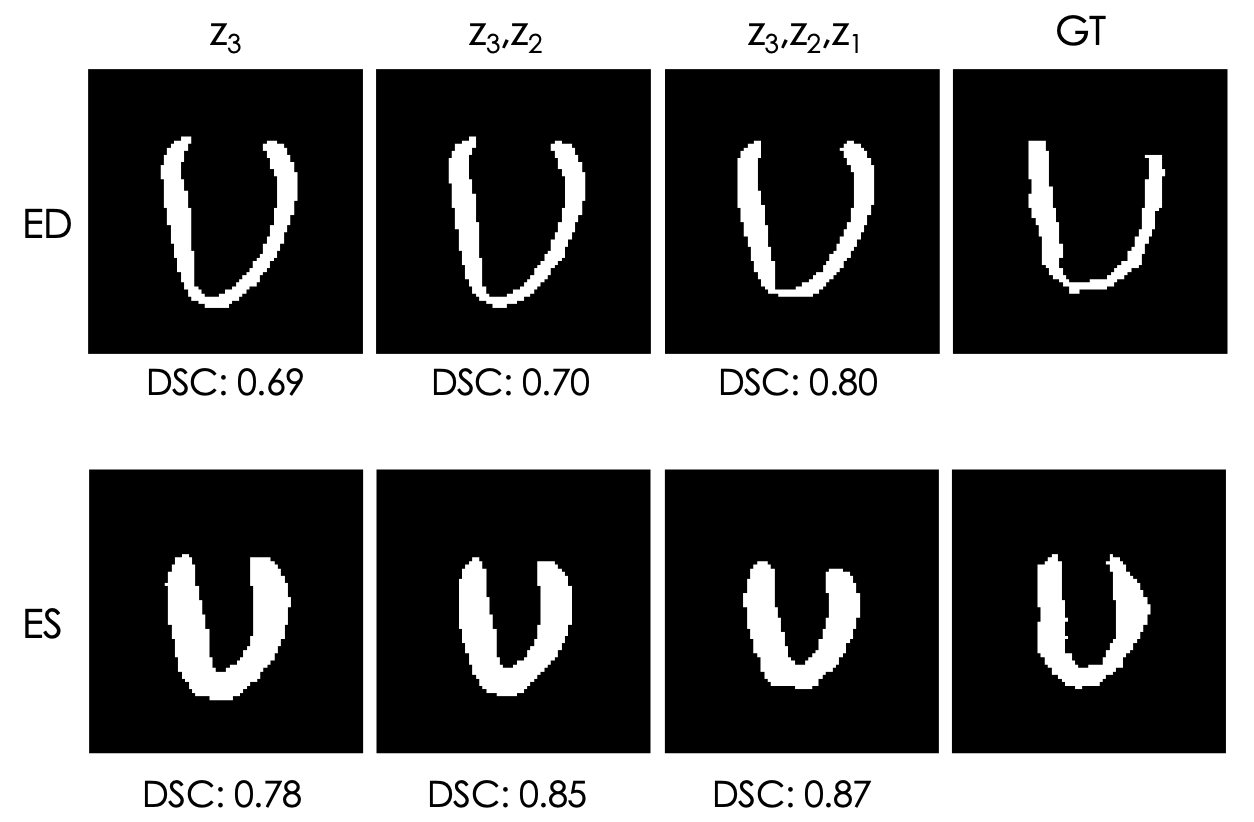}
\caption{Long-axis section of an HCM patient reconstructed segmentations at ED and ES by the LVAE+MLP model using only $z_3$ information (first column) or also using the posterior information of the other latent spaces $(z_2, z_1)$ .  Last column: ground-truth (GT) segmentation. DSC = Dice Score between the segmentation at that column and the GT.}
\label{fig:recHVOL}
\end{figure}
\newpage
\begin{figure}[h!]
\centering
\includegraphics[width=0.85\columnwidth]{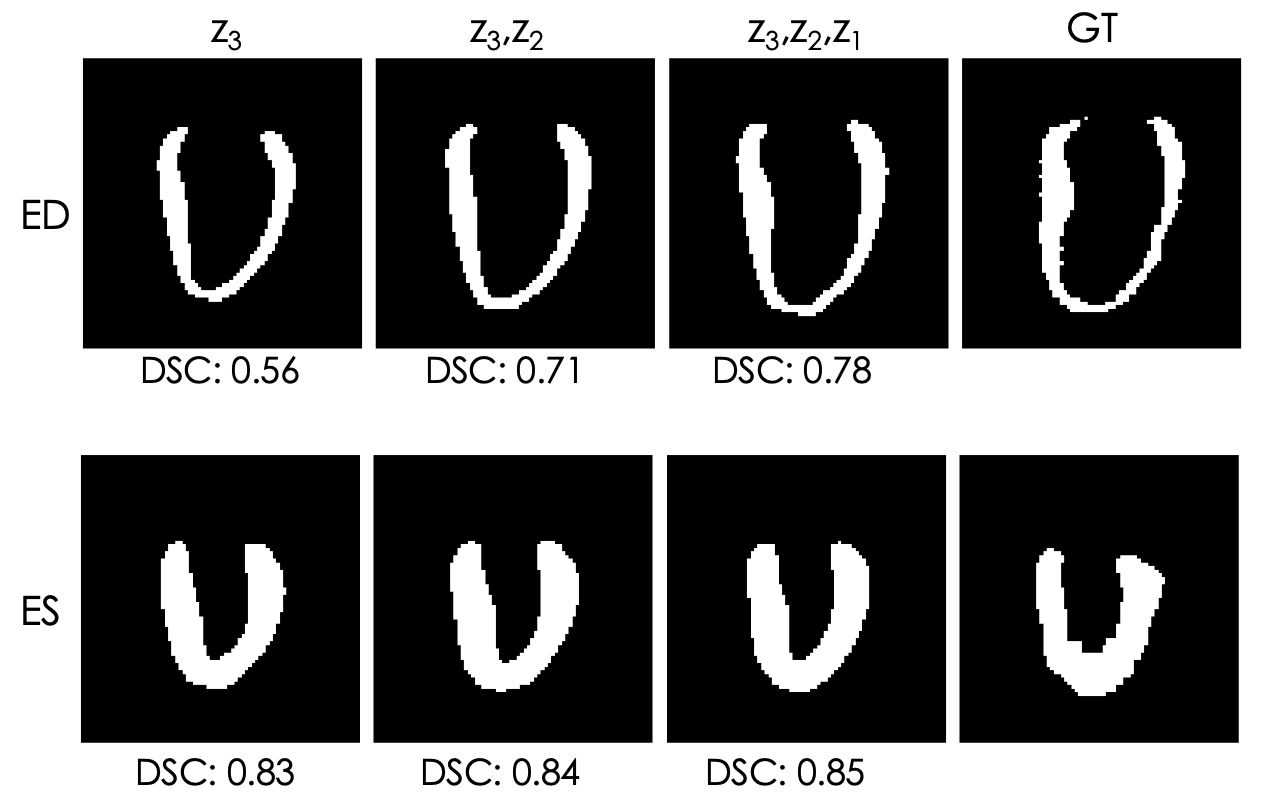}
\caption{Long-axis section of an HCM patient reconstructed segmentations at ED and ES by the LVAE+MLP model using only $z_3$ information (first column) or also using the posterior information of the other latent spaces $(z_2, z_1)$ .  Last column: ground-truth (GT) segmentation. DSC = Dice Score between the segmentation at that column and the GT.}
\label{fig:recHVOL}
\end{figure}

\newpage

\section*{Supplementary Data 5 - Brain Application - LVAE+MLP Archicture}
\begin{figure}[h!]
\centering
\includegraphics[scale=0.23]{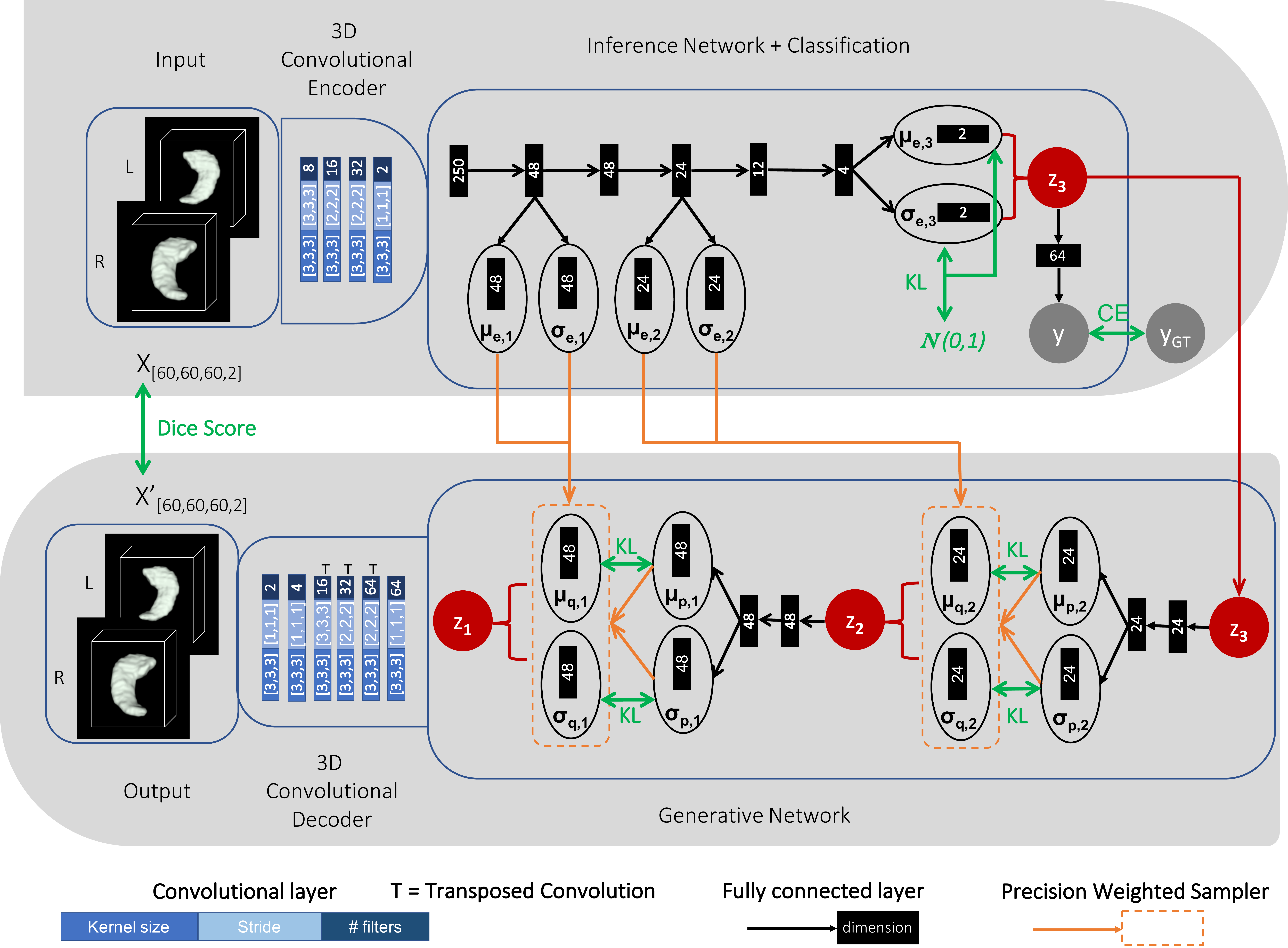}
\caption{Detailed scheme of the LVAE+MLP architecture adopted in this work for AD vs healthy controls experiments. Top: encoder model; Bottom: decoder model. The green arrows indicate the loss function terms used to train the network.}
\label{fig:LVAEarchi}
\end{figure}

\newpage

\section*{Supplementary Data 6 - Cardiac Application - Standard Imaging Measures} 

\begin{table}[h!]
\centering
\begin{tabular}{l c c c c c}
\hline
 & HCM &  & Hvol &    \\
\hline
 & Mean & SD & Mean & SD \\
Age at recruitment / first CMR & 54.87 & 15.97 & 37.51 & 12.94 \\
Females (\%) & 27.1 &  & 36.8 &  \\
BSA (m$^3$) & 1.91 & 0.23 & 1.80 & 0.19 \\
Left ventricular end-diastolic volume (mm$^3$) & 134.77 & 36.35 & 141.77 & 30.52 \\
Left ventricular end-systolic volume (mm$^3$) & 35.28 & 17.43 & 48.53 & 14.51 \\
Left ventricular ejection fraction (\%) & 74.45 & 9.74 & 66.12 & 5.04 \\
Left ventricular mass (g) & 182.00 & 64.84 & 109.84 & 32.29 \\
Max wall thickness (mm) & 18.33 & 4.85 & 7.37 & 3.52\\
\hline
\end{tabular}
\caption{Table of population characteristics for the cardiac dataset. Information for 34 HCMs patients were not available. The total number of healthy volunters (HVols) subjects is 451, total number of HCMs is 402.}
\end{table}
\end{document}